\documentclass{aa}

\usepackage{siunitx}
\usepackage{scalerel}
\usepackage{graphicx}
\usepackage{txfonts}
\usepackage{natbib}
\usepackage{xcolor}
\usepackage{multirow}
\usepackage{float}
\usepackage{tikz}
\usetikzlibrary{svg.path}
\usepackage{booktabs}
\usepackage[normalem]{ulem}

\usepackage{hyperref}

\usepackage{orcidlink}

\begin{document}

   \title{A statistical study of lopsided galaxies using random forests}

   \author{Valentina Fontirroig\inst{1}{\orcidlink{0009-0000-1589-9766}}
          \and
          Facundo A. G\'omez\inst{1}{\orcidlink{0000-0003-4232-8584}}
          \and
          Marcelo Jaque Arancibia\inst{1}{\orcidlink{0000-0002-8086-5746}} 
          \and 
          Arianna Dolfi\inst{1}{\orcidlink{0000-0002-0704-7798}}
          \and 
          Nicol\'as Monsalves\inst{1}{\orcidlink{0000-0002-4129-8195}}
          }

   \institute{$^{1}$ Departamento de Astronom\'ia, Universidad de La Serena,
              av. Raul Bitr\'an, La Serena, Chile\\
              \email{valentina.fontirroig@userena.cl}
             }


\abstract
  {Lopsided galaxies are late-type galaxies that feature a non-axisymmetric  {disk} caused by an uneven distribution of their stellar mass or light. Despite being a relatively common perturbation, there are still several questions regarding its origin and the information that can be extracted from them about the evolutionary history of late-type galaxies. Previous observational and numerical studies have suggested a correlation between lopsidedness and galaxy assembly history and internal structure.}
   {The advent of several large  {multiband} photometric surveys will allow us to statistically analyze this perturbation, with information that was not previously available. This paper aims to develop a method  {of} rapidly and automatically  {pre-classifying} late-type galaxies  {as} lopsided and symmetric, purely based on the galaxies' internal parameters. This method allows us to test the hypothesis that lopsidedness is a strong indicator of peculiar internal galaxy properties, rather than an indicator of the present-day environment they are hosted in.}
   {We  {selected} a sample of $\approx 8,000$ late type galaxies from the Illustris TNG50 simulation. A Fourier  {decomposition} of their stellar mass surface density  {was} used to label galaxies as lopsided and symmetric. We trained a  {random forest} binary classifier to rapidly and automatically identify this type of  {perturbation}, exclusively using galaxies' internal properties. We  {explored} different  {algorithms} to deal with the imbalance nature of our data, and  {selected} the most suitable approach based on the considered metrics.}
   {We show that our trained algorithm can provide a very accurate and rapid pre-classification of lopsided galaxies. The excellent results obtained  by our classifier, trained with features that do not account for the galaxies environment, strongly  {support} the hypothesis that lopsidedness is mainly a tracer of galaxies' internal structures. We also show that similar results can be obtained when considering as input features observable quantities that are readily obtainable from  {multiband} photometric surveys.}
   {Our results show that algorithms such as  {the ones} considered allow  {for the} rapid and accurate pre-classification of lopsided galaxies from large  {multiband} photometric surveys, allowing us to explore whether lopsidedness in present-day  {disk} galaxies is connected to  {galaxies'} specific evolutionary histories.} 
   

\maketitle

\section{Introduction}

Lopsided galaxies feature a non-axisymmetric  {disk}, caused by an uneven distribution of stellar mass or light. Observational studies have shown that up to $30\%$ of nearby galaxies display  {a} certain degree of lopsidedness \citep{zarrix1997,rudnick&rix1998,bournaud2005,vaneymeren2011}. The term  {``lopsidedness''} was first coined by \cite{baldwin1980} to refer to those galaxies in their sample  {for which there was} a strong asymmetry in the HI gas density distribution between  {the} opposite sides.  {As was} discussed in \citet[][and references therein]{jog}, this asymmetry can significantly impact the dynamical structure and evolution of the host galaxy, causing enhanced  {star-forming} regions, fueling the central active galactic nucleus, and redistributing matter, among  {other matters}.

Despite lopsided galaxies being an ubiquitous object in the nearby  {Universe}, this asymmetry has received less attention than other commonly studied perturbations \citep[e.g.,][]{sellwood2013, conselice2014,erwin2019}.  Moreover, the origin of this asymmetry is not quite well understood, as both galaxies in the field and in denser environments present lopsidedness. Different mechanisms have been proposed as the main driver of this asymmetry, such as asymmetric gas accretion \citep{phookun1993, bournaud2005}, tidal encounters \citep{weinberg1994,rudnick2000, gomez2016}, satellite accretion \citep{walker1996, zarrix1997},  {the} response of the  {disk} to the distorted dark matter halo \citep{jog1997,jog2002}, and off-centered  {disk} \citep{noordermeer2001}, among others.  

Interestingly, several works have  {found differences in the structural properties of lopsided galaxies with respect to the ones of} more symmetrical late-type galaxies. In particular, \cite{reichard2008} studied a sample of 25,155  {low-redshift} ($z < 0.06$) galaxies in the Sloan Digital Sky Survey \citep[SDSS,][]{sdss}, and showed that lopsided galaxies tend to have  {a} lower concentration and stellar mass density within their half light radius than symmetrical galaxies. This suggests that there is a correlation between lopsidedness and the structural properties of the galaxies. Using a sample of late-type galaxies from the Illustris TNG50 simulation, \cite{silvio} found a similar strong correlation between lopsidedness and the internal properties of galaxies. Specifically, they found an anticorrelation between lopsidedness and the tidal force exerted by the inner regions on the outskirts of their galactic  {disk}. This result indicates that less gravitationally cohesive  {disk} galaxies are more susceptible to  {developing} this asymmetry when exposed to external perturbations. {Indeed, these results, based on numerical models, are in agreement with \citet{1999ApJ...522..661J, 2000ApJ...542..216J}, who studied the self-consistent response of an axisymmetric galactic  {disk} perturbed by a constant lopsided halo potential. They showed that the  {disk} self-gravity plays a crucial role in the determination of the net lopsided distribution in the  {disk}.} \cite{arianna} extended this study by considering a larger sample of $z=0$ TNG50  {disk}-like galaxies, located in different environments. They showed that, independently of the environment, while symmetric galaxies are typically assembled at early times ($\sim 8 - 6$ Gyr ago), with a relatively short and intense burst of central star  formation, lopsided galaxies assembled over a longer time periods, with less prominent initial bursts and a subsequent milder and constant star formation rate up to $z = 0$.

 {Large current and upcoming} observational surveys, such as S-Plus \citep{splus}, J-Plus\citep{jplus}, J-PAS\citep{jpas}, and LSST\citep{lsst}, will enable  {us} to identify and characterize lopsidedness in a very large number of well-resolved galaxies in the local Universe. This will be crucial to further study the connection between such perturbation and the  {internal galaxy} properties, and to test current model predictions and understand the origin of lopsidedness considering their star formation history in relation with the environment. However, as the volume of data increases, using traditional approaches to study and characterize this non-axisymmetry, such as visual inspection \citep[e.g.,][]{baldwin1980, richter1994},  {the} identification of surface brightness residuals with respect to unperturbed distributions \citep[e.g.,][]{conselice2000}, and Fourier decomposition \citep[e.g.,][]{zarrix1997, reichard2008}, can become a limiting task. All these techniques require human supervision and intervention (such as visual inspection),  {and thus} result in cumbersome and slow approach to study lopsidedness in larger volumes of data, which could also result in missing important information or discoveries.

\begin{table*}
    \caption{List of the internal parameters of the 7,919  {disk}-like galaxies obtained from TNG50-1.}             
    \label{tableparam}      
    \centering          
    \begin{tabular}{l c c}     
    \hline     
    \noalign{\smallskip}
    Parameters & Symbol & Description\\ 
    \noalign{\smallskip} 
    \hline          
    \rule{0pt}{4ex}  
       Tidal Parameter$^1$ & $T_{\rm P}$ &  Represents the tidal force applied by the inner galaxy regions ($R<R_{50}$) to \\
      & & the materials located at distances equal to $R_{90}$. Defined as $T_P =M_{50}/R_{90}^3$. \\  
      
    \rule{0pt}{4ex}  
       Central Stellar Mass Density & $\mu_*$ & Density of the stellar mass contained inside $R_{50}$. Defined as $\mu_{star}=M^{*}_{50}/\pi R_{50}^2$.\\
       & &  Here $M^{*}_{50}$ represents the stellar mass within $R_{50}$. \\
    \rule{0pt}{4ex}  
       Minor-to-major axis & \textit{c/a} & Ratio between the minor axis \textit{c} and major axis \textit{a}. Describes the\\
       & & shape of the inner galactic regions.\\
    \rule{0pt}{4ex}  
        {Disk}-to-total mass & \textit{D/T} & Ratio between the  {disk's} mass and the total mass of the galaxy. Used to select\\
       & &  {disk}-like galaxies.\\
    \rule{0pt}{4ex}  
       Concentration & \textit{C} &  Ratio between $R_{90}$ and the effective radius $R_{50}$. Defined as $C=R_{90}/R_{50}$. \\
    \rule{0pt}{4ex}  
       Effective Radius & $R_{50}$ & Radius of the galaxy containing $50\%$ of the stellar mass.\\
    \rule{0pt}{4ex}  
        {Disk} Extension & $R_{\rm ext}$ & Defined as 1.4 times the radius of the galaxy containing $90\%$ of\\
        & & the stellar mass ($R_{90}$).  \\
    \rule{0pt}{4ex}  
       Half-mass & $M_{50}$ & Total (baryonic and dark matter) mass of the galaxy enclosed in $R_{50}$.\\
    \rule{0pt}{4ex}  
       Star Formation Rate & \textit{SFR} & Total stellar mass created from gas and dust, per year. \\
    \rule{0pt}{4ex}  
       Spin Parameter$^1$ & $\Lambda(R)$ & Galactic  {disk} stellar spin, which is a proxy of the apparent stellar angular\\ 
       & &  momentum. Calculated within the inner stellar half-mass radius.  \\
    \rule{0pt}{2ex} \\
       
    \hline               
    \end{tabular}
    \tablebib{(1) \citet{lagos2017}.}
    
\end{table*}

To avoid this problem, machine learning algorithms are a helpful tool used to automate and speed up the classification of objects such as galaxies. Different algorithms can be used for different tasks depending on the problem to solve. In particular, some methods are defined as supervised, as they require a subsample of already labeled data to train and test a model, generally referred as training set. A few of these algorithms consist of ensemble methods,  {such as random forests} \citep{randomforestpaper} and  {gradient boosting} \citep{friedman2001}, artificial neural networks,  {such as convolutional neural networks \citep{oshea2015} and recurrent neural networks \citep{rumelhart1987}}, distance-based algorithms,  {such as nearest-neighbor algorithms} \citep{cover1967}, among others. Some examples of their application are morphological classification of galaxies \citep{ball2004, dieleman2015, huertas2015,farias2020}, classification of variable stars using time-domain \citep{aguirre2019,monsalves2024}, and the estimation of photometric redshifts \citep{Zhang2013,lee2021}. On the other hand, unsupervised algorithms such as clustering,  {such as} k-means \citep{lloyd1982} and  {hierarchical} clustering \citep{guha2000}, and dimensional reduction algorithms,  {for example principal component analysis} \citep{jolliffe2002}, are trained with unlabeled data. They are typically employed for anomaly detection \citep{baron2017,giles2019, sarkar2022}, feature selection \citep{ZHENG2008} and extraction \citep{wang2017}, and even galaxy classification \citep{hocking2018}.

Given the  {strong} previously reported correlation between lopsidedness and the structural properties of galaxies, this paper aims to use using machine learning techniques to automatically classify galaxies  {as} lopsided and symmetric by only using their internal properties. Our goal is to develop a machine learning classifier algorithm designed as a fast  {preselection} tool rather than a replacement for other  {methods} that provide crucial information such as a Fourier decomposition. We also seek to explore whether an accurate classification of this asymmetry can be obtained without including any direct information regarding the environment inhabited by the galaxies. Our selected machine learning algorithms were trained and tested over a large sample of galaxies obtained from the IllustrisTNG simulation. We  {also determined} the key parameters that  {enable} the correct classification of lopsided galaxies. The organization of this paper is as follows. In Sect. \ref{data} we list the selection criteria to obtain the internal and observational parameters of disk-like galaxies extracted from the TNG50-1 simulation. In Sect. \ref{methods} we describe the procedure we follow to implement the classification algorithms. In Sect. \ref{results} we list and analyze our results. The conclusions and discussion are finally presented in Sect. \ref{conc}.

\section{Data}
\label{data}

In this section we present the criteria to select the necessary dataset to train and test our selected pre-classification models, discussed in Sect. \ref{sectmodels}. In particular, we use galaxy models extracted from the fully cosmological simulation, Illustris TNG50 \citep[][]{2019MNRAS.490.3234N,2019MNRAS.490.3196P}. For each galaxy model, we compute internal parameters that are commonly measured in observational studies to classify galaxies' morphology.

\subsection{The IllustrisTNG project}

IllustrisTNG (hereafter TNG), successor of the Illustris project \citep{genel2014, vog2014, nelson2015}, is a set of cosmological, gravo-magnetohydrodynamical simulation  {run} with the moving-mesh code \textsc{arepo} \citep{springel2010}. IllustisTNG builds upon its predecessor model \citep{genel2014} by incorporating an updated physical model \citep{Pillepich2018} which accounts for stellar evolution, gas cooling, feedback and growth from supermassive blackholes, among others. In particular, the improved model for the feedback of the low accretion mode in  {supermassive} black holes resulted in a reduction of the discrepancies with observational constraints identified in the original Illustris simulations, such as the galaxy color bimodality \citep{Nelson2018}. These improvements make IllustrisTNG a powerful tool for comparisons with observational data.

TNG consists of three simulations with different volumes: ~50$^3\rm Mpc$, ~100$^3\rm Mpc$, and ~300$^3\rm Mpc$, referred as TNG50, TNG100, and TNG300, respectively. Each simulation was run with different mass and spatial resolution. As a result, the three realizations complement each other. The largest simulation box, TNG300, enables the study of galaxy clustering and provides the largest statistical galaxy sample. On the other hand, TNG50 provides the smallest galaxy sample at the high mass end, but it has the highest mass resolution overall. Therefore, it enables a more detailed look at the morphology of galaxies and its structural properties. TNG-100 falls somewhere in between these two other simulations.

In this work, due to its mass and spatial resolution, we make use of the publicly available TNG50-1 model \citep{pillepich2019,nelson2019b}. Having a dark matter and baryonic mass resolution of $4.5\times 10^5\rm M_{\odot}$ and 8.5$\times 10^4\rm M_{\odot}$, respectively, TNG50-1 allow us to resolve the structure of $10^9 \rm  M_{\odot}$ stellar  {disk} with at least $10^4$ stellar particles, enabling a better characterization of their morphology \citep{nelson2019a}. 

The cosmological model adopted in IllustrisTNG is a flat $\Lambda$CDM Universe with the following parameters: Hubble constant H$_0=67.8\rm{kms}^{-1}\rm{Mpc}^{-1}$, total matter density $\Omega_m = 0.3089$, dark energy density $\Omega_{\Lambda} = 0.6911$, baryonic matter density $\Omega_b = 0.0486$, rms of mass fluctuations at a scale of 8 $h^{-1}\rm{Mpc}$ $\sigma_8=0.8159$, and a primordial spectral index $n_s=0.9667$ \citep{planck}.

\subsection{Selection criteria}
\label{crit}

We focus our study on central and satellite  {disk}-like galaxies, identified within the redshift range \textit{z=\rm 0} to \textit{z=\rm 0.5}. The \textit{z} range considered allow us to obtain a large number of galaxy models to train our classification algorithm. Note that, even though a given galaxy will be present at different snapshots of the simulation, their detailed structure will evolve \citep[see e.g.,][]{silvio} and, thus, it will serve as input for the training process.

Following \cite{arianna} based on our selection criteria, we consider galaxies with:

\begin{itemize}
    \item $N_{\rm tot,stars} \geq 10^4$, where $N_{\rm tot,stars}$ represents the number of bound stellar particles. This is used to make sure that galaxies have enough stellar particles to be reasonably well resolved. Considering that the baryonic mass resolution is $\sim 10^5\rm M_{\odot}$, as mentioned before, the minimum stellar mass considered is $\gtrsim 10^9 \rm M_{\odot}$.

    \item $f_{\rm e} > 0.4$, where $f_{\rm e}$ represents the circularity fraction, defined as the fractional mass of the stellar particles with circularity $\epsilon > 0.7$. The latter has previously shown to reliably select orbits confined to a  {disk} \citep{aumer2013}. $f_e$ kinematically quantifies the  {disk's} shape, thus ensuring that the galaxies selected are considered  {``disky''} \citep{joshi2020}.

    \item $R_{90}\geq 3\rm kpc$. This ensures that the structure of the galactic  {disk} is clearly resolved. 
    
\end{itemize}

\begin{figure*}
    \centering
    \resizebox{0.95\hsize}{!}
        {\includegraphics{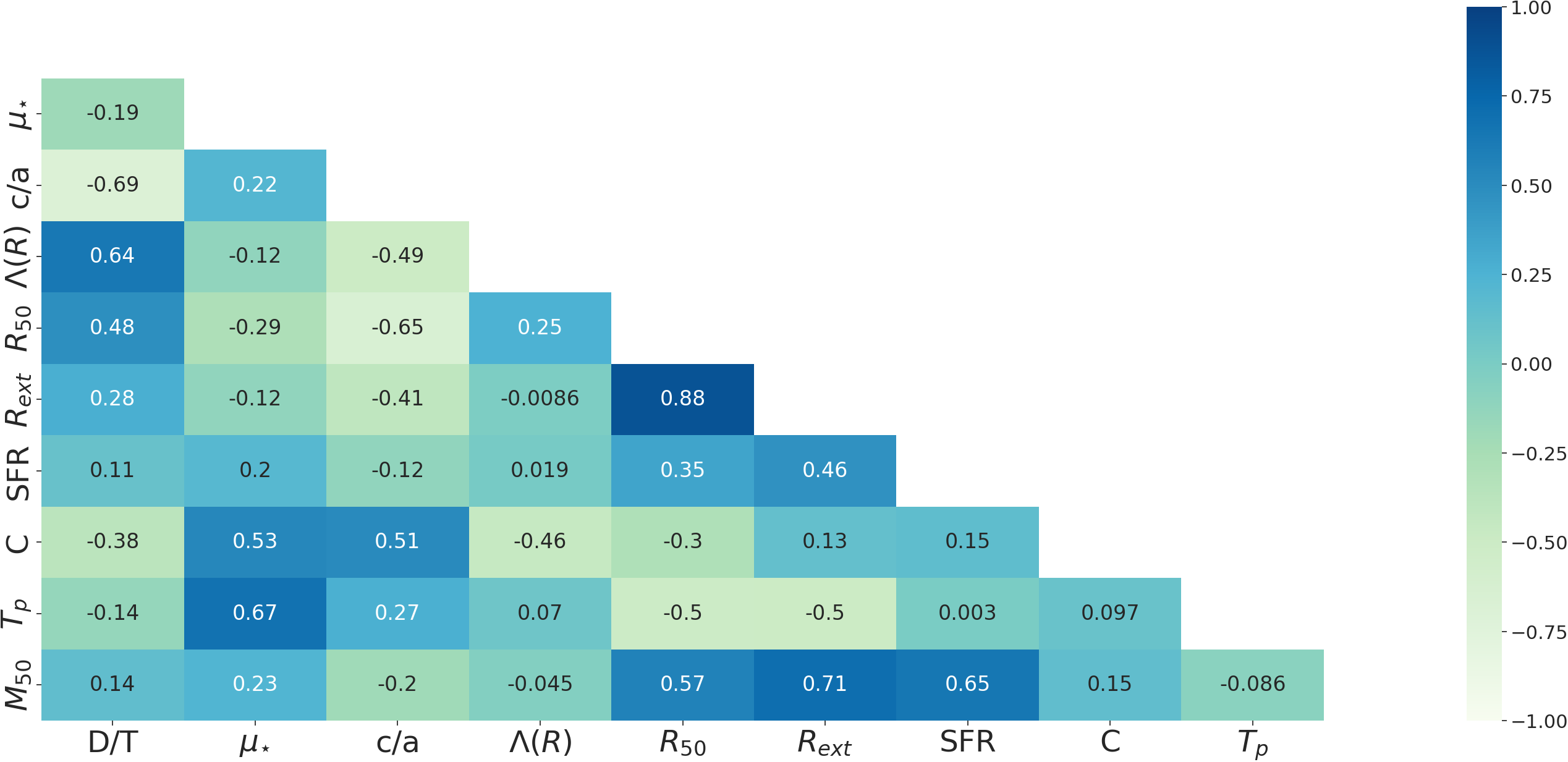}
        }
    \caption{Heatmap of the Pearson Coefficient Correlation of the features listed in Table \ref{tableparam}.}
    \label{corr_feat}
\end{figure*}

These criteria result in a sample of 7,919 late-type galaxies. In Table \ref{tableparam} we list the parameters measured from each galaxy that are later used to train our  {pre-}classification models. These parameters are computed as described in \citet{arianna}, and references there in. Note that all selected parameters characterize galaxies internal properties and do not explicitly account for the environment in which the galaxies are located. Moreover, previous works have shown that some of these parameters, such as the  {disk} central stellar density, $\mu_*$, and its extension, $R_{\rm ext}$, are expected to be strongly linked to the occurrence of lopsided perturbations. In Fig. \ref{corr_feat} we quantify the Pearson Correlation Coefficient between the listed parameters. Checking the parameters' correlation is an important first step to ensure an accurate representation of the classifier's results, as having highly correlated data (Pearson correlation values of 1 and -1) can lead to a misinterpretation of the importance of some parameters. In our case, we note that our parameters do not show a strong correlation, with the exception of $R_{50}$ and $R_{\rm ext}$, which have a score of 0.88. This suggests that there is no issue in applying all the selected parameters in our classifier.

\section{Methods}
\label{methods}

In this work we make use of the  {random forest} \citep[hereafter RF;][]{randomforestpaper} algorithm and its variations to study our selected dataset. Since we deal with a supervised algorithm, it is necessary to count with a training and testing set where galaxies are already labeled as lopsided or symmetric galaxies. A Fourier  {decomposition} of the light/mass distribution is often used to quantify asymmetries \citep{zarrix1997,reichard2008,silvio,arianna}. We  {used} the radial distribution of the $m=1$ mode to label our dataset. To prepare our data before applying it to the models, we  {partitioned} the dataset into a training set and a testing set comprising $70\%$ and $30\%$ of the total sample, respectively. To do so, we  {employed} \textsc{stratifiedshufflesplit} from \textsc{scikit-learn}\footnote{\url{https://scikit-learn.org}}.

In this section we first discuss how lopsidedness is measured in our models, and then introduce and summarize the main characteristics of  {random forests} and its variations. We also discuss our particular application and the metrics used to measure its performance.

\subsection{Measuring lopsidedness}
\label{meas_lop}

\begin{figure*}
    \centering
    \resizebox{0.85\hsize}{!}
        {\includegraphics{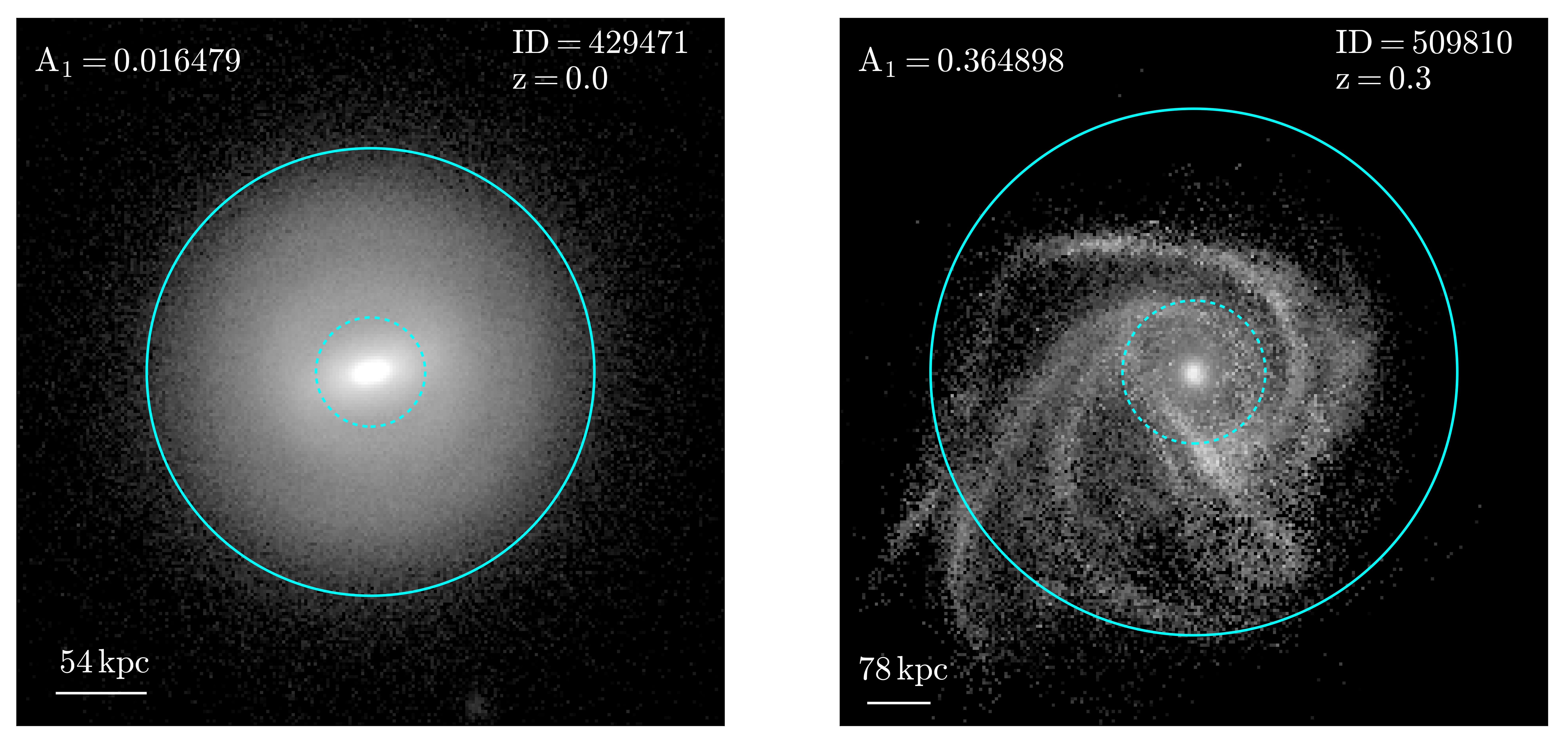}}
    \caption{V-band face-on projected surface brightness distribution of a symmetric \textit{(left)} and lopsided \textit{(right)} galaxy, considered as examples of the classification made by $A_1$. Their respective $A_1$ value, ID (as in TNG50-1), and redshift snapshot are plotted on the upper side. On the lower left, the box size considered for each galaxy is also plotted. For both images, the dashed cyan line represents the radius $R_{50}$ and the solid cyan line represents the radius 1.4$R_{90}$, which are the limits of the radial interval used in the Fourier decomposition.}
    \label{gal_examples}
\end{figure*}

\begin{figure}
 \centering
    \includegraphics[width=0.9\linewidth]{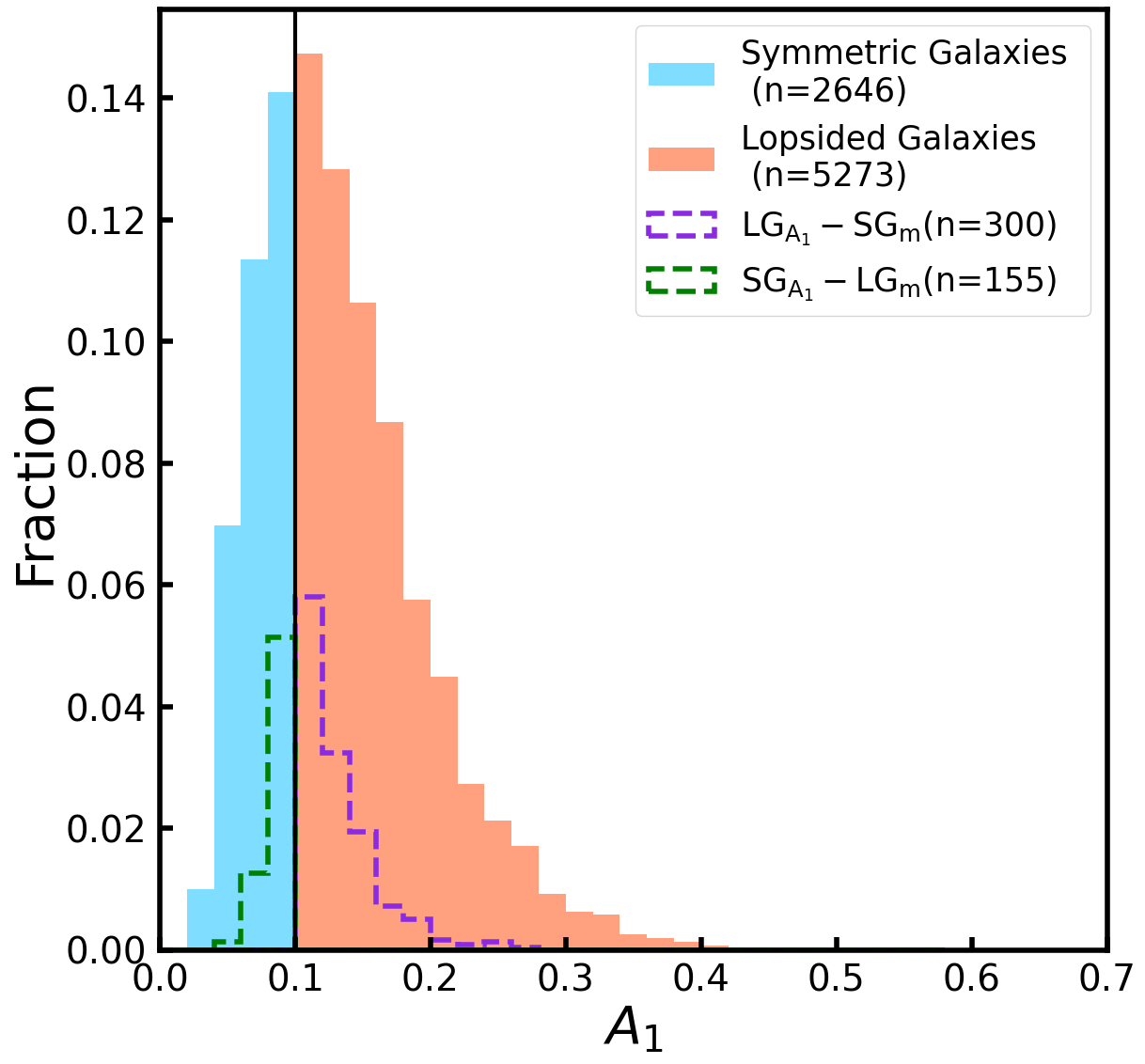}
       \caption{$A_1$ distribution of our total sample. $A_1$ is defined as the averaged strength of the $m=1$ mode of the Fourier  {decomposition} for each stellar particle within the radial range  $\rm R_{50} -1.4R_{90}$. The black line represents the threshold used to distinguish between lopsided (orange) and symmetric galaxies (blue). The dashed distributions represent the incorrect classifications of the galaxies made by SMOTE+RF for testing set, as we discuss later in Sect. \ref{results}. The  {dashed purple} distribution represents the actual symmetric galaxies classified by the model as lopsided galaxies $\rm (SG_{A_1}-LG_m)$, and the  {dashed green} distribution represents the actual lopsided galaxies classified by the model as symmetric galaxies $\rm (LG_{A_1}-SG_m)$. Each distribution has in parenthesis their respective number.}
          \label{a1dist}
\end{figure}

\begin{figure*}
    \centering
    \resizebox{0.9\hsize}{!}
        {\includegraphics{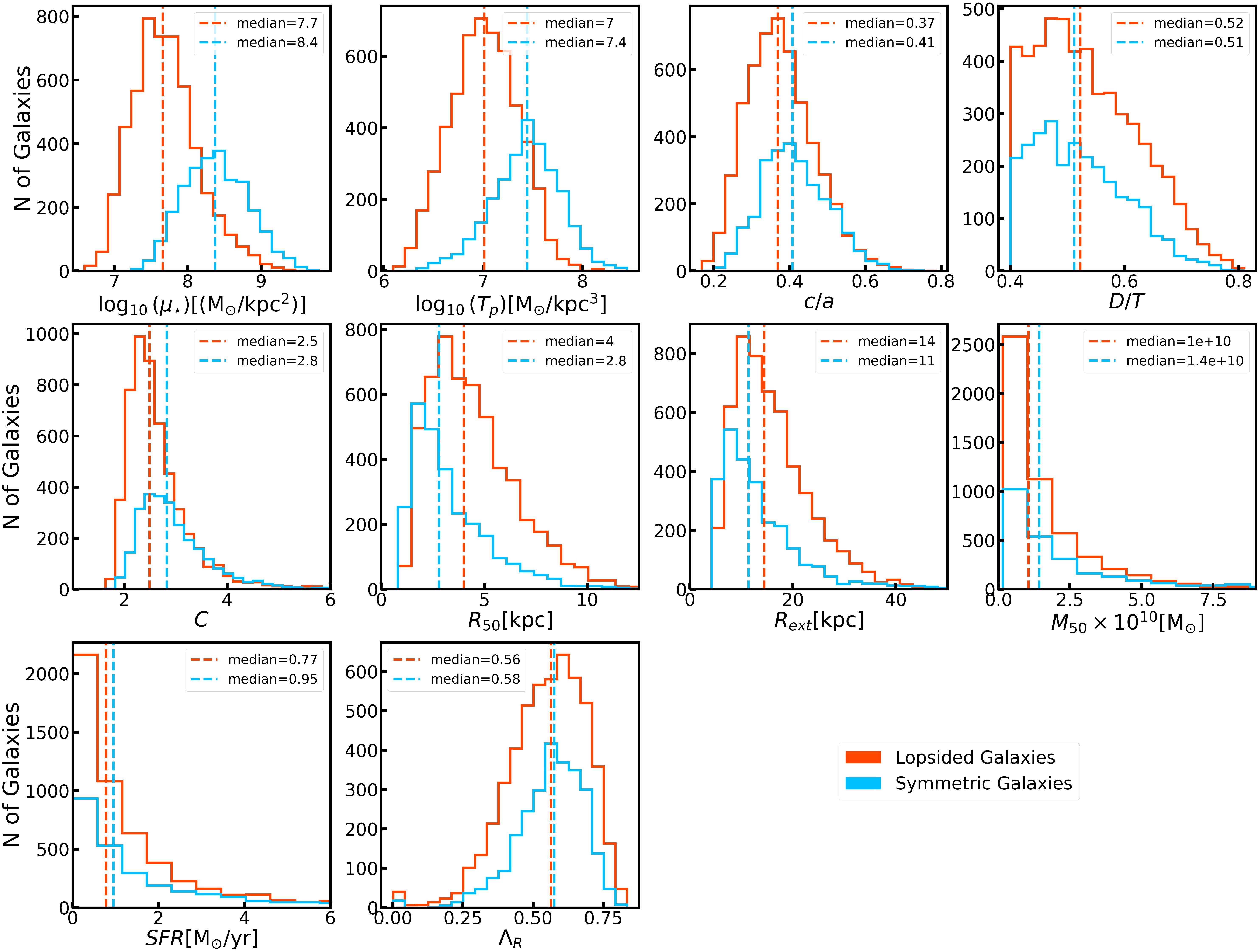}}
    \caption{Distribution of  parameters selected to characterize our galaxy sample. These parameters are used as features by the  {random forest} classifier. The orange and blue distributions represent lopsided and symmetric galaxies, respectively. The  {dashed colored} lines represent their respective median.}
    \label{feat_dis}
\end{figure*}

To label the galaxies in our sample between lopsided and symmetric, we  {applied} a Fourier  {decomposition}. To do so, we measure the amplitude of the first mode $m=1$ of the stellar  {disk} density distribution, $A_1$, which quantifies the asymmetry of the stellar mass distribution. Before doing so, we  {took} into account a few considerations. First, it is crucial to ensure that each galaxy is projected face-on, as the Fourier  {decomposition} is highly sensitive to the  {disk} inclination. To do so, we  {rotated} each galaxy such as the  {z axis was} aligned with the  {disk} angular momentum vector. Secondly, to focus our analysis on stellar  {disks}, we consider only stellar participles located within a cylinder of width equal to $1.4R_{90}$, and a height equal to $2h_{90}$. Here, $h_{90}$ is defined as the vertical distance above and below the  {disk} plane enclosing $90\%$ of the total galaxy stellar mass. The adopted definition for the  {disk} extent  {allows} us to reach their outer regions without introducing contamination from the stellar halo. We have tested several definitions for the  {disk} extent, and found that, overall, the results are not significantly affected by our definition.

The Fourier decomposition for the stellar mass distribution  {was} calculated as follows:

\begin{equation*}
    C_m(R_j,t) = \sum_i M_i e^{(-im\phi_i)},
\end{equation*}

where $M_i$ and $\phi_i$ are the mass and the azimuthal coordinate of the i-th stellar particle,  {respectively}. The $A_1$ radial profile is then calculated as follows:

\begin{equation*}
    A_1(R_j,t) = \frac{B_1(R_j,t)}{B_0(R_j,t)},
\end{equation*}

where $B_1(R_j,t)$ and $B_0(R_j,t)$ are the amplitude, or strength, of the $m=1$ and $m=0$ mode, respectively, within a certain radius {, $R_j$, and a certain snapshot, \textit{t}}. In general, the amplitude of the Fourier decomposition is given by:

\begin{equation*}
    B_m(R_j,t) = \sqrt{a_m^2(R_j,t) + b_m^2(R_j,t)},
\end{equation*}

where $a_m(R_j,t)$ and $b_m(R_j,t)$ are defined as the real and imaginary values of $C_m(R_j,t)$ for the m-th mode, respectively.

The averaged value of $A_1(R,t)$ (hereafter $A_1$) at a given time, $t$, and over a certain radial interval is then used as the global or large-scale lopsidedness indicator. In general, if $A_1 > 0.1$, the galaxy is considered lopsided. For values of $A_1 < 0.1$ galaxies are considered symmetric. This threshold has been widely adopted in the literature, where both large observational and simulated galaxies were considered \citep[e.g.,][]{jog, reichard2008, silvio, arianna}. The radial interval considered to calculate the global $A_1$ parameter has varied between different works. For instance, \cite{zarrix1997} studied the lopsidedness distribution of a sample of 60 field spiral galaxies, using the radial interval of $\rm (1.5-2.5)$  {disk} scale lengths. On the other hand, \cite{reichard2008} measured the lopsidedness of a sample obtained from SDSS in the radial interval $R_{50}$-$R_{90}$. \citet{vaneymeren2011} reached distances up to 4 to 5  {disk} scale lengths to study the asymmetries of the  {disks}' outer regions. In our case, we use $\rm R_{50}-1.4R_{90}$, as we find that this radial interval best represent the non-axisymmetry of the sample. 

As an example of the classification made by $A_1$, Fig. \ref{gal_examples} shows the face-on projections of the surface brightness distribution in the V-band of two clearly classified cases. Here the dashed and cyan lines indicate the lower and upper radial limits, respectively, considered to compute $A_1$. Considering their respective $A_1$ values, the galaxy on the left is classified as a strong symmetric example with ~$A_1=0.02$, while the galaxy on the right is classified as a strong lopsided example with a value of ~$A_1=0.36$.

The resulting $A_1$ distribution of our sample is shown in Fig. \ref{a1dist}. The light blue and orange shaded areas indicate the distribution for symmetric  and lopsided classified galaxies, based on the selected $A_1$ threshold (black line). Notably, our sample is imbalanced;  {meaning that} we have a higher quantity of lopsided galaxies with respect to the symmetric cases.
Out of the total sample size of 7,919 galaxies, 5,273 (i.e., $\sim 65\%$) are classified as lopsided, while 2,646 (i.e., $\sim 35\%$) as symmetric. We note that we find a larger fraction of lopsided galaxies than observations in the local Universe (e.g., $\sim 30\%$; \citealt{zarrix1997,reichard2008}). As previously discussed in \cite{arianna}, this difference can be likely attributed to the different radial interval used to measure the global lopsidedness $A_1$. For this reason, we are finding a larger fraction of lopsided galaxies than observations, due to the fact that we are reaching out to larger galactocentric radii where the lopsided amplitude is stronger (see also \citealt{silvio}). The resulting imbalance imposes a great challenge for the training and testing of our selected machine learning algorithms. In the following section, we dive deeper into this issue and describe the methods we use to address it.

Lastly, Fig. \ref{feat_dis} shows the distribution of our selected parameters, subdividing both types of galaxies to stress their differences. The dashed lines indicate the median of the corresponding distributions. The first and second top panels show the distributions of $\mu_{*}$ and $T_{\rm P}$. It is evident that the two galaxy types show the largest differences in these two parameters. As expected, lopsided galaxies typically show significantly smaller $\mu_{*}$ than their symmetric counterparts. Similarly, lopsided galaxies exhibit smaller values of $T_{\rm P}$. This trends are in agreement with previous results \citep{reichard2008,zaritsky2013,silvio} that highlighted that both types of galaxies are indeed characterized by different internal structures.

\subsection{Classification algorithms}
\label{sectmodels}

 {RF is} a type of supervised algorithm that  {poses} a great advantage in the automation of different classification and regression tasks. For instance, it can describe different complexity relations between the parameters, or features, of a sample considering their assigned label. It can also works with a wide variety of different datasets and sizes, among other advantages. In the case of astronomy, it is clear that the use of machine learning algorithms, such as  {random forests}, have grown as a result of the exponential increase in data with the current and next-generation surveys and telescopes. As application examples,  {random forests pose} a great alternative to classify different sources in different wavelengths \citep{gao2009}, estimation of photometric redshifts \citep{carliles2010}, perform automatic classification of light curves of variable stars \citep{alerce}, predict underlying gas conditions of the circumgalactic medium \citep{appleby2023}, identify galaxy mergers \citep{guzman2023} and estimate different galaxies' physical properties \citep{mucesh2021}, among other applications.

 {RF} consists of an ensemble, or collection, of decision trees. A decision tree is a tree-like predictive model composed of nodes, where the sample is recursively divided by conditions in the form of $ x_i^{(j)}< \boldsymbol{X}_j$, the latter being the j-th feature and $x_i^{(j)}$ a certain threshold based on the j-th feature. In other words, decision trees divide the input space, which depends on the selected feature/s by the decision tree, to create subspaces that are able to differentiate between the different classes. The quality of this division is measured by the  {``purity''} of the subspace, where the purer it is, the more datapoints from the same class are assigned. The final nodes, called leaves or terminal nodes, result  {from} either fully partitioning the sample or until all leaves have less than the minimum quantity to split a node, which is determined by a certain parameter in the tuning process, discussed below. For classification tasks, these final nodes provide predictions of a certain class based on the resulting probability or, in the case of regression tasks, they provide a numeric value. This process is done firstly on a learning set, or training set, where then a new unseen dataset is propagated over the tree to predict the corresponding class or numeric value.

Although decision trees have numerous advantages due to their intrinsic nature (e.g., they can be used by any kind of sample, they have an easy hyperparameter customization or tuning, and they also estimate the feature importance aside from class predictions), they are easy to overfit. This means that a decision tree may be less accurate when predicting unseen data during testing, as the model tends to overly fit to the training set. RF avoids this issue by training non-correlated decision trees, each on a subsample with replacement of the training set, thus reducing the variance while maintaining high accuracy. For a binary classification task, which is our focus, each decision tree classifies the data as either positive or negative class. Then, the final prediction of the RF is the class predicted by more than half of the trees. This method is called bagging or {bootstrap aggregating} \citep{bagging}.

Due to our dataset being imbalanced, as seen in Fig. \ref{a1dist}, using a RF classifier could lead to an inaccurate classification. The training and testing of {random forests} are performed considering bootstrapped samples of the corresponding data sets. As each sample follows the same distribution as the original dataset, the majority class would have more predictions in favor, thus having more accurate results than the minority class. To avoid this issue affecting our results, we employ two different algorithms. The first one consists on oversampling the minority class of the training set and then apply it to a RF classifier. To do the oversampling, we use \textsc{smote} \citep{smote} from \textsc{imbalanced-learn}\footnote{\url{https://imbalanced-learn.org}}. This method creates new  {``synthetic''} data by interpolation between two close datapoints in the multidimensional feature space; in our case, a 10 dimensional feature space. The second algorithm consists of using Balanced Random Forests \citep[hereafter BRF;][]{brf}, where we use the \textsc{balancedrandomforestclassifier} method from \textsc{imbalanced-learn}. In this case, the bootstrapped sample is only considered for the minority class, whereas the majority class is randomly sampled with replacement, matching the size of the minority class. This avoids manually oversampling the dataset and it is directly performed by each decision tree.

\begin{table}
    \caption{Results of the hyperparameter tuning using \textsc{randomizedsearchcv} for each model, SMOTE+RF and BRF.}            
    \label{hypres}      
    \centering                          
    \begin{tabular}{l c c }      
        \hline              
        \noalign{\smallskip}
        Hyperparameters & SMOTE+RF & BRF \\ 
        \noalign{\smallskip}      
        \hline                        
        \noalign{\smallskip} 
          number of trees & 1500 & 128  \\      
          minimum samples split & 5 & 5     \\
          minimum sample leaf & 2 & 2      \\
          maximum features  & sqrt & log2    \\
          maximum depth  & 75 & 200     \\ 
          sampling strategy & - & all \\
          \noalign{\smallskip} 
        \hline                                  
    
    \end{tabular}
\end{table}

To have an optimal performance of both classifiers using our datasets, we perform an hyperparameter tuning. This involves finding the best combination of parameters from the models to yield the best results. The hyperparameters involved in the fitting of the RF classifiers are listed in Table \ref{hypres} with their respective results. To tune both models, we use \textsc{randomizedsearchcv} with number of iterations $n_{iter}=10$ and, as cross-validation, \textsc{RepeatedStratifiedKFold} with number of repeats $n_{repeat}=10$ and number of splits $n_{splits}=5$. For SMOTE+RF and BRF, we use the default values of $n_{iter}$, $n_{repeat}$, and $n_{splits}$. To avoid unnecessary complexity in the calculations, we retain the default values for the current and following analysis. For the tuning process, we select a range of possible values for each hyperparameter and then apply it to the randomized search. This generates random combinations of hyperparameters and selects the combination that yields the best performance based on a chosen metric, which in our case is balanced accuracy. It is worth highlighting the significant difference in the number of trees between both classifiers, where SMOTE+RF has 1,500 trees in comparison with BRF, which only has 128. This discrepancy in the number of trees might be attributed to the added complexity and variability introduced to the minority class by the SMOTE oversampling process. As it creates synthetic data, the complexity and variability of the sample increases, requiring SMOTE+RF to utilize a larger ensemble of trees to effectively generalize the data and achieve robust results.

\subsection{Evaluation metrics}
\label{metrics}

To measure the performance of both SMOTE+RF and BRF, we  {used} the following metrics considering the use of binary classifiers:

\begin{itemize}
    \item Precision: Ratio of the number of correctly predicted positive class to the total number of predicted positive class. Expressed as:
    
        \begin{equation*}
            {\rm Precision} = \frac{\rm TP}{\rm FP+TP},
            \label{eqprec}
        \end{equation*}

    {where TP represents  {true positives} ( {i.e.,} symmetric galaxies with $A_1<0.1$ pre-classified as symmetric by our selected models) and FP represents  {false positives} ( {i.e.,} lopsided galaxies with $A_1 > 0.1$ pre-classified as symmetric).}\\

    \item TPR: Ratio of the number of correctly predicted positive class to the number of actual positive class. Expressed as:

        \begin{equation*}
            {\rm TPR} = \frac{\rm TP}{\rm FN+TP},
            \label{eqrec}
        \end{equation*}

    {where FN represents  {false negatives} ( {i.e.,} symmetric galaxies with $A_1<0.1$ pre-classified as lopsided by our models).}\\

    \item F1-score: Harmonic mean of precision and TPR. Expressed as:

      \begin{equation*}
        {\rm F1-score} = 2 \times \frac{\rm precision \times TPR}{\rm precision + TPR}
        \label{eqf1}
        \end{equation*}

    \item  {True negative rate} (TNR) or specificity: Ratio of the correctly predicted negative class to the total number of the actual negative class. Expressed as:
    
        \begin{equation*}
            {\rm TNR} = \frac{\rm TN}{\rm{FP+TN}},
            \label{eqtnr}
        \end{equation*}

     {where TN represents  {true negatives} ( {i.e.,} lopsided galaxies with $A_1>0.1$ pre-classified as lopsided by our models).}\\

    \item Balanced  {accuracy}: 
    Average of the recall obtained for each class. Expressed as:
    
        \begin{equation*}
    \label  {bacc}
            {\rm balanced~accuracy} = \frac{1}{2}\left(\rm TNR \times TPR \right)
        \end{equation*}

    \item Geometric  {mean} (G-mean): Square root of TNR and \textit{TPR}. Expressed as:

        \begin{equation*}
        {\rm G-mean} = (\rm TNR \times TPR)^{1/2}
        \label{eq-gmean}
        \end{equation*}

    \item ROC-AUC: Calculates the area under the Receiver Operating Characteristic (ROC) curve, by using the trapezoidal rule, which approximates the area under the curve as a series of trapezoids. Considering a series of points in the ROC curve, in the form of $(x_1,y_i),(x_2,y_2),...,(x_N,y_N$), the area under the curve is expressed as:

        \begin{equation*}
            {\rm ROC-AUC} = \sum_{i=1}^{N-1} \frac{(x_{i+1}-x_i)(y_i+y_{i+1})}{2}
        \end{equation*}
        
\end{itemize}

The selected metrics are used to evaluate the results of our classifiers. In particular, precision, TPR, and F1-score are important metrics to evaluate the performance of any type of model. However, these metrics are all sensitive to imbalanced dataset. As a result, they could mislead the algorithm during the training and validation process. To avoid this, we focus the analysis of our classifiers to TNR, balanced accuracy, and G-mean. These metrics are selected following \cite{brf} work, which ensure a correct analysis due to the imbalanced nature of our dataset. Lastly, we also consider ROC-AUC for the analysis, as it gives us an important insight on how the model is performing without any effect of the imbalance. Still, we present the values for TPR, precision, and F1-score, as a reference.

\section{Results and analysis}
\label{results}

In this section we introduce and analyze the results of the  {RF} algorithm trained to pre-classified subsamples of lopsided and symmetric galaxies. The quantification of its accuracy was performed by contrasting against the galaxies' $A_1$ values on each pre-classified subsample. We then explore in detail the most common caused behind misclassifications.

\subsection{Classification results}
\label{RF_sect}

\begin{table}
    \caption{Metric scores of the selected model, SMOTE+RF, applied to the testing set.}                
    \centering                          
    \begin{tabular}{ l c c }      
        \hline              
        \noalign{\smallskip} 
        &
        \multicolumn{1}{c}{SMOTE+RF} &
        \multicolumn{1}{c}{BRF}\\ 
        \noalign{\smallskip} 
        \cline{2-3}
        \noalign{\smallskip} 
         Metric & Score & Score\\
        \hline      
        \noalign{\smallskip} 
         Precision  & 0.702$\pm$0.013 & 0.675$\pm$0.007\\
         TPR & 0.797$\pm$0.019 & 0.833$\pm$0.014\\
         F1-score & 0.746$\pm$0.012 & 0.746$\pm$0.007\\
         ROC-AUC & 0.813$\pm$0.010 & 0.816$\pm$0.006 \\
         TNR  & 0.830$\pm$0.011 & 0.799$\pm$0.00\\
         G-mean & 0.813$\pm$0.010 & 0.816$\pm$0.006\\
         Balanced-Accuracy & 0.813$\pm$0.010 & 0.816$\pm$0.006\\
        \noalign{\smallskip} 
        \hline
    \end{tabular}
    \label{metric_scores}
\end{table}

As a brief  outline of our classification pipeline, we train the classifiers mentioned in Sect. \ref{sectmodels} with 5,542 galaxies, constituting $\sim 70\%$ of the total sample. This enables the algorithm to obtain important underlying patterns and/or relationships between the galaxies and their features, which are then used for the prediction in the final step. The remaining galaxies are consider for the testing set, which compromises a total of 2,377 galaxies, or $30\%$ of the remaining sample. For each galaxy, these decision trees produce a class prediction, either lopsided or symmetric. The class that is predicted in more than half of the decision trees is taken as the final prediction for that galaxy. Due to the imbalanced nature of our dataset, we define lopsided as the negative class and symmetric galaxies as the positive class. Usually, the majority class is better represented and naturally favorable by the algorithm over the minority class. To avoid this problem, we designate the minority class as the positive class, which helps with the interpretability of metrics, such as TPR, precision, and ROC-AUC for rare cases. Since we also obtain a proxy of the probability of a galaxy being in the positive/negative class, we test different thresholds, or cut-offs, to classify the samples and to explore how  {such a} threshold can affect our results. As a default, this threshold is set at 0.5.  {Galaxies with probabilities equal or greater than 0.5} are labeled as the positive class or, in our case, symmetric galaxies. Galaxies with probabilities  {lower} than this value are labeled as the negative class,  {in our case being} lopsided galaxies. Our analysis showed that differences in the results obtained between the different cut-offs is negligible. Therefore, the following analysis was performed with the default value, 0.5, for SMOTE+RF and BRF. We emphasize that the  {random forest} classifier is designed as a fast pre-selection tool rather than a replacement for a Fourier  {decomposition, and} thus we consider the results given by our models as a  {``preselection'' or ``preclassification''.}

The results of each model's performance for the testing set are listed in Table \ref{metric_scores}. Each value of the metrics is obtained by averaging the result scores of each iteration of a cross-validation with number of iterations $n_{iter}=10$ and taking into consideration its standard deviation. It is clear that both classifiers provide similar results, with comparable values in most metrics. Based on this, we select as our classifier SMOTE+RF since it results in a better TNR metric. As previously discussed, we are working with an imbalanced data set, with more than $70\%$ of the data belonging to the negative class (lopsided objects). Thus, a high TNR indicates a better performance for the most populated class of our sample.

\begin{figure}
    \centering
    \includegraphics[width=0.9\linewidth]{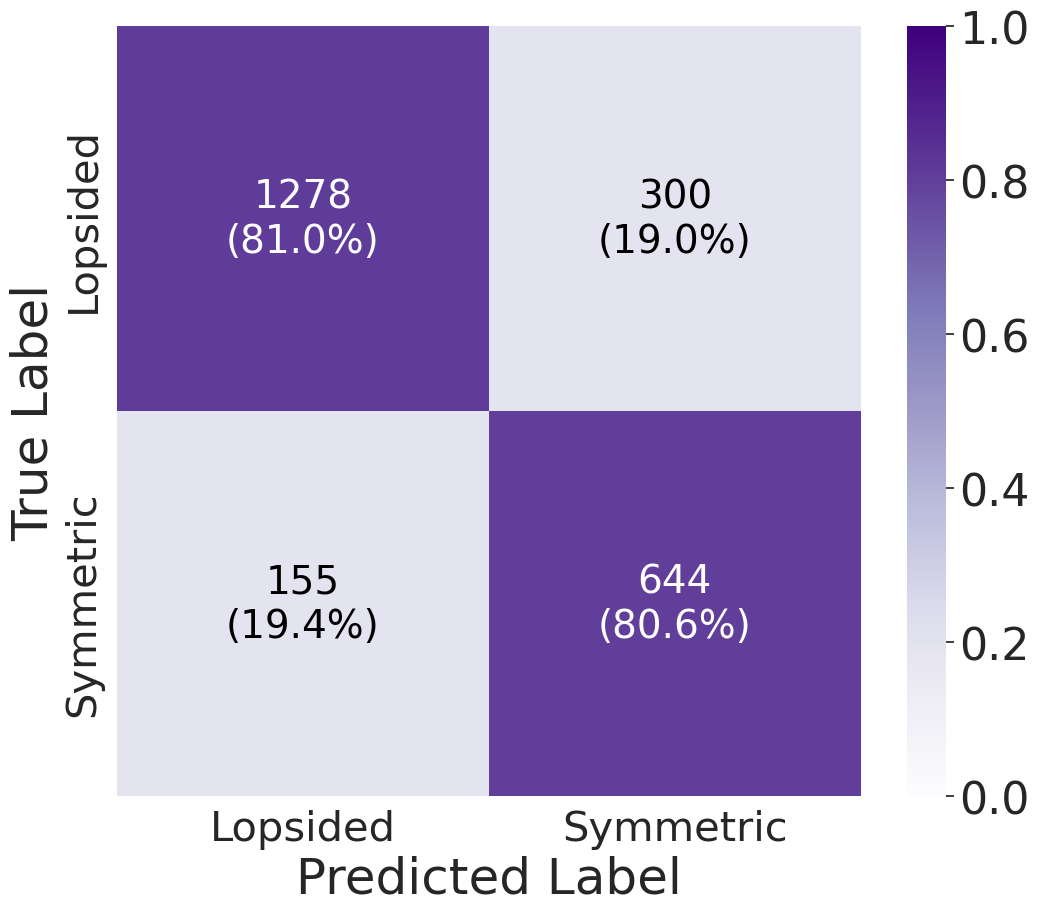}
    \caption{Confusion matrix for the testing set of the best model, SMOTE+RF. The  {x axis} is the predicted class or predicted label, and the  {y axis} is the actual class or actual label. The percentage with respect each type of galaxy set is on parenthesis.}
    \label{cm}
\end{figure}

Fig. \ref{cm} shows the confusion matrix (CM) for SMOTE+RF. The  {x axis} indicates the predicted class or label, obtained from the classifier, and the  {y axis shows} the actual class or label, obtained from the $A_1$ parameter. In general, a CM allows us to visually inspect the fractions of correct and incorrect classification we have obtained. In our testing sample, and based on our $A_1$ classification criteria, we count with 1,578 true lopsided and a total of 799 true symmetric galaxies. Interestingly, our classifier is able to correctly  {pre-classify} $81\%$
of the lopsided objects and approximately the same amount for their symmetric counterparts. In absolute number, we obtain a total of 1,922 correctly  {pre-classified} galaxies, against 455 incorrectly  {pre-classified} objects. It is worth highlighting the very good performance of the SMOTE+RF classifier, which has been purely obtained based on features that are related to our simulated galaxies internal properties. No information about environments has been introduced during the training process.

\subsection{Interpretation of the random forest classification}

\begin{figure}
 \centering
    \includegraphics[width=0.9\linewidth]{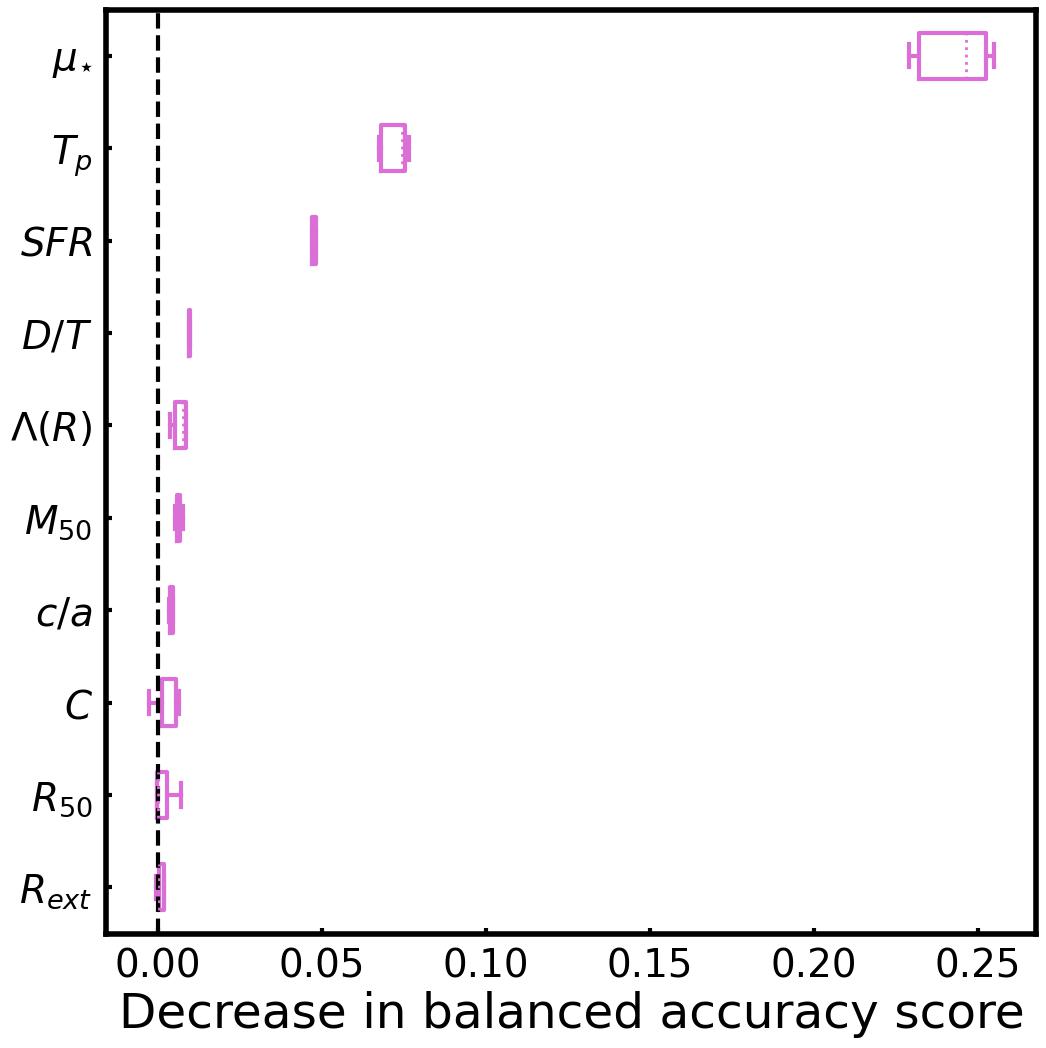}
    \caption{Box plot of each feature from the testing set, ranked by their importance as determined by the feature permutation attribute from SMOTE+RF. Each box represents the range of the different scores obtained from a cross-validation with $n_{iter}=5$. The inner dashed line represents the median value of each distribution. The whiskers on each box represent the minimum and maximum value of each distribution.}
    \label{boxplot}
\end{figure}

\begin{figure*}
    \centering
    \resizebox{0.8\hsize}{!}
        {\includegraphics{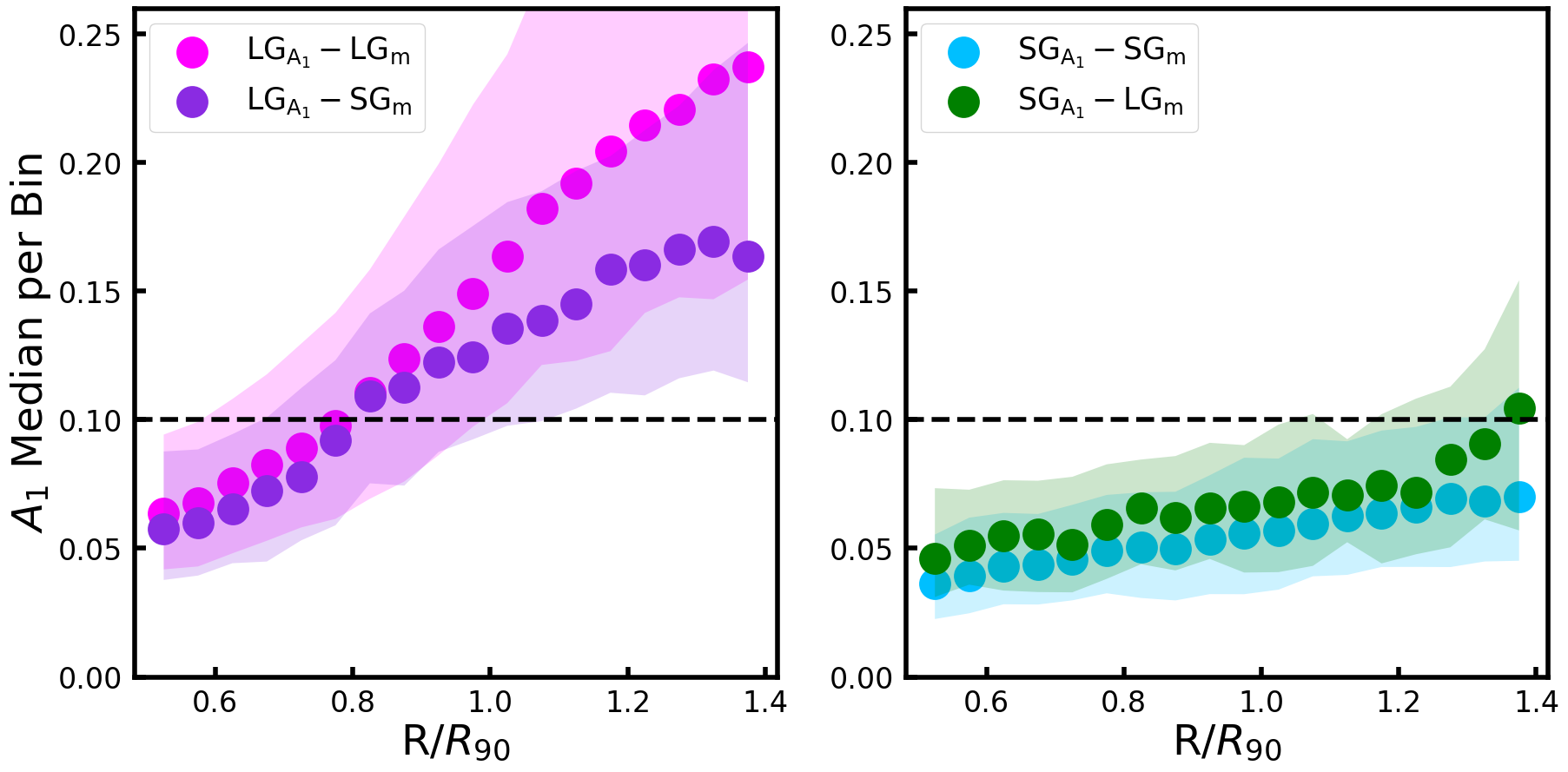}}
    \caption{Radial profiles of $A_1$ for our four classification cases, calculated as the median of $A_1$ for each bin with respect to $R_{90}$. The fuchsia and blue distributions represent the correctly classified lopsided galaxies $\rm (LG_{A_1}-LG_m)$ and symmetric galaxies $\rm (SG_{A_1}-SG_m)$, respectively. The green distribution represents  {symmetric galaxies pre-classified as lopsided }$\rm (SG_{A_1}-LG_m)$ and the purple distribution represents  {lopsided galaxies pre-classified as symmetric} $\rm (LG_{A_1}-SG_m)$. The shaded areas represent the 25th and 75th percentiles of each sample.}
    \label{a1vsropt}
\end{figure*}

Supervised algorithms, including  {RF algorithms}, suffer from interpretability of the decisions leading to the classification. This is often called the ``black box'' problem. In  {RF}, it arises due to the high quantity of decision trees added to the ensemble. In this section we interpret and analyze the decisions lead by the model to pre-select the galaxies between lopsided and symmetric by ranking the importance of the features used in the classification process.

We  {used} the \textsc{permutation\_importance\_} attribute from \textsc{randomforestclassifier}. There are various methods  {of} ranking feature importance, but given the continuous nature of our dataset, where no categorical features are used for training or testing, we rely solely on \textsc{permutation\_importance\_}. This attribute works by permuting, or shuffling, the values of each feature and calculating the resulting decrease  {in} a specified metric, which by default is accuracy, defined as the fraction or count of the correct predictions. The decrease in the score is then used to rank each feature: the higher the score, the more it affects the model's performance, thus making the feature important for the model to maintain a higher accuracy. However, since our dataset is imbalanced, using accuracy would not return an accurate representation of the importance of our features. To address this issue, we use balanced-accuracy instead. As discussed in Sec. \ref{metrics}, this metric represents the averaged fraction of correct classified galaxies for both the negative and positive class. In this, each class contribute equally to the final score, regardless of its size. Considering that the accuracy metric disproportionately favors the majority class in imbalanced datasets due to its overrepresentation, balanced accuracy is great alternative to avoid inaccurate results.

The results of this procedure are shown in Table \ref{feat_imp}, where it lists the rank of each feature obtained by \textsc{permutation\_importance\_}. Considering that we want to focus on the performance of SMOTE+RF with unseen data, we only calculate the feature importance for the testing set. We obtain each score by averaging the iterations of a cross-validation with $n_{iter}=5$ and taking into consideration its standard deviation. This analysis clearly shows that both $\mu_*$ and $T_{\rm P}$ are the highest-ranked parameters, with $\mu_*$ ranked first and $T_{\rm P}$ ranked second. As a way to better visualize this, Fig. \ref{boxplot} also shows the variation of balanced-accuracy with a box plot. Each box represents the distribution of the score value for each iteration. The dotted line inside each box is the median of the distribution, and each whisker represents the first and last score value. Indeed, we note that $\mu_*$ is the top-ranked parameter overall, indicating that it is the most important parameter to consider in the classification process made by SMOTE+RF. As we previously mentioned, and as seen in Fig. \ref{feat_dis}, lopsided and symmetric galaxies are characterized by different $\mu_*$ distributions. This is in agreement with previous results \citep{reichard2008,zaritsky2013,silvio,arianna}, where lopsided galaxies tend to show significantly lower a  densities in the inner regions (as defined by their $R_{50}$) with respect to the symmetric counterparts. 

\begin{table}
    \caption{Feature Importance of each parameter calculated by \textsc{permutation\_importance\_} for SMOTE+RF.}                
    \centering                          
    \begin{tabular}{c  l c }      
        \hline              
        \noalign{\smallskip}
         Rank &  Feature & Score \\
        \hline      
        \noalign{\smallskip} 
            1& $\mu_*$&0.242930$\pm$0.010621\\
            2& $T_{\rm P}$ & 0.072285$\pm$0.003824\\
            3& \textit{SFR} & 0.047444$\pm$0.005302\\
            4& \textit{DT}&0.009005$\pm$0.002047\\
            5& $\Lambda(R)$&0.006586$\pm$0.002012\\
            6& $M_{50}$&0.006234$\pm$0.000864\\
            7& \textit{c/a}&0.004205$\pm$0.000934\\
            8& \textit{C}&0.003042$\pm$0.003434\\
            9& $R_{50}$&0.001760$\pm$0.002815\\
            10&$R_{\rm ext}$&0.000721$\pm$0.000990\\
        \noalign{\smallskip} 
        \hline
    \end{tabular}
    \label{feat_imp}
\end{table}

Although not as important as $\mu_*$, $T_{\rm P}$ and \textit{SFR} also play an important role in the classification process in comparison with the rest of the features. This is also in agreement with previous results, where an  {(anti)} correlation between lopsidedness and $T_{\rm P}$ \citep[e.g.,][]{gomez2016} and a correlation between lopsidedness and \textit{SFR} \citep[e.g.,][]{conselice2000} have been reported. In particular, $T_{\rm P}$ represents a proxy of the tidal force exerted by the inner galactic regions on the outer  {disk} material. In other words, it indicates how gravitationally cohesive a galaxy is. The relevance of this parameter is clearly reflected in the separation between the distribution of both types of galaxies, as previously seen in Fig. \ref{feat_dis}, where lopsided galaxies tend to have lower values of $T_{\rm P}$ than symmetric galaxies. These findings align with the conclusions of \cite{silvio} and \cite{arianna}, which propose that lopsided perturbations serve as indicators of intrinsic galaxy properties, rather than being predominantly driven by environmental processes. In other words, galaxies with low central stellar densities are weakly gravitationally cohesive and, thus, are more susceptible to lopsided perturbations, independently of a particular perturbing agent. On the other hand, \textit{SFR} ranking third place is an interesting result, as it is been shown that there is a correlation between $A_1$ and current \textit{SFR} \citep{zarrix1997,rudnick2000,reichard2009}. As discussed by \cite{Lokas_2022}, some internal properties of lopsided and symmetric galaxies can be linked with their current \textit{SFR},  {such as} lopsided galaxies having bluer colors, larger gas fractions, and lower metallicity than symmetric galaxies. Moreover, \cite{arianna} showed that lopsided galaxies tend to be, on average, significantly more star forming than symmetric galaxies at later times. Symmetric galaxies, on the contrary, have an earlier assembly with shorter and more intense star forming bursts. As a result, and considering galaxies with similar stellar masses at the present-day, while symmetric galaxies tend to develop a more  {pronounced} central region at earlier times, lopsided galaxies tend to form a larger fraction of their stellar populations later, typically developing a more extended stellar  {disk} and less dense inner regions. Lastly, Fig. \ref{boxplot} shows the relative importance of the remaining 7 features. It is clear that they have a minimal impact on the classification procedure.

To analyze the classification made by SMOTE+RF, we plot in Fig. \ref{a1vsropt} the median of $A_1$ as a function of radius for the four cases defined by the classifier. To generate this figure, the radial extension of each simulated galaxy was normalized by its corresponding $R_{90}$. We focus on the radial interval $\rm (0.5-1.4)R_{90}$, as it is the considered interval for the Fourier decomposition. The shaded areas represent the 25th and 75th percentiles of the distribution. In the left plot, the fuchsia distribution represents  {true lopsided galaxies pre-classified by our model as lopsided}, $\rm (LG_{A_1}-LG_m)$, whereas the purple distribution represent  {true lopsided galaxies pre-classified by our model as symmetric}, $\rm (LG_{A_1}-SG_m)$. The right plot is the same as the left one but for symmetric galaxies. Here, the cyan color represent the distribution of {true symmetric galaxies pre-classified by our model as symmetric}, $\rm (SG_{A_1}-SG_m)$, while in green we show {true symmetric galaxies pre-classified by our model as lopsided}, $\rm (SG_{A_1}-LG_m)$. We note that both incorrectly classified cases,  {$\rm (LG_{A_1}-SG_m)$ and $\rm (SG_{A_1}-LG_m)$}, do not follow the same trend as the correctly classified distributions. In the case of $\rm (LG_{A_1}-SG_m$) on the left plot, from 0.5$R_{90}$ to 0.9$R_{90}$ the magnitude of $A_1$ starts increasing at the same rate than the correctly classified sample. However, from 0.9$R_{90}$ onward, the slope is less steep, meaning that the magnitude of $A_1$ does not increase as much as in $\rm (LG_{A_1}-LG_m$). In other words, while incorrectly classified galaxies have indeed an outer perturbed region, the strength of the perturbations is typically weaker with respect  to correctly classified galaxies. On the right panel we show that the $A_1$  profile of both correctly $\rm (SG_{A_1}-SG_m)$ and incorrectly classified galaxies $\rm (SG_{A_1}-LG_m)$ remains below the 0.1 threshold chosen to classify lopsided galaxies based on the $A_1$ parameter. Nonetheless, incorrectly classified symmetric galaxies tend to have larger $A_1$ values at all radii and they do cross the threshold at the outermost edge. In the following section we explore in detail the main reasons that drove the SMOTE+RF method to misclassify these galaxies.

\begin{figure*}
    \centering
    \resizebox{0.9\hsize}{!}
        {\includegraphics{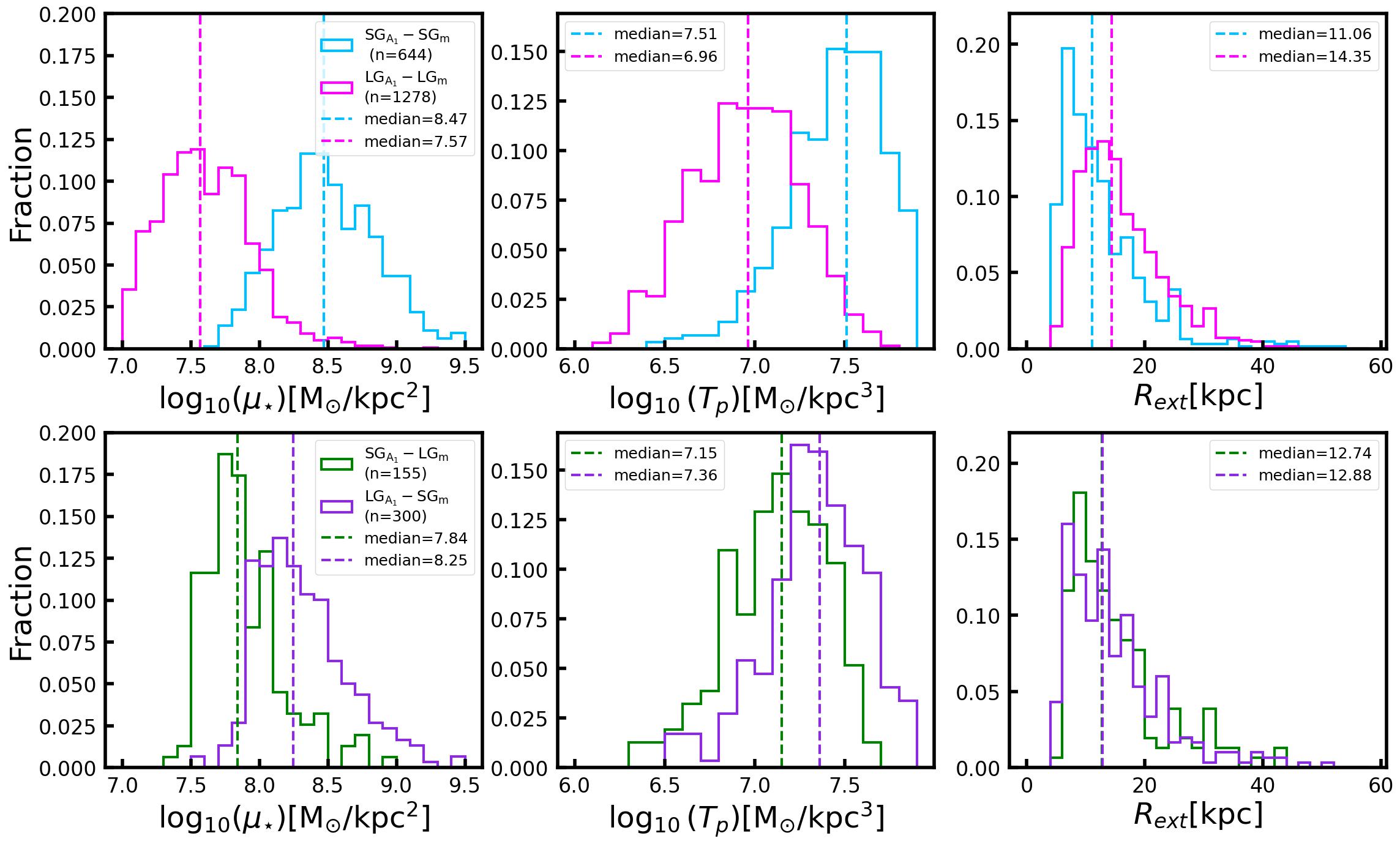} 
        }
    \caption{Normalized distribution of $\mu_*$ \textit{(left)}, $T_{\rm P}$ \textit{(middle)}, and $R_{ext}$ \textit{(right)}, considering the correct \textit{(upper)} and incorrect \textit{(bottom)} classification made by SMOTE+RF. Each distribution has been normalized by their corresponding number of galaxies of each subsample. Their respective number of galaxies is in parenthesis. The fuchsia and blue distributions represent the correctly classified lopsided galaxies $\rm (LG_{A_1}-LG_m)$ and symmetric galaxies $\rm (SG_{A_1}-SG_m)$, respectively. The green distribution represents symmetric galaxies  {pre-classified} as lopsided $\rm (SG_{A_1}-LG_m)$ and the purple distribution represents lopsided galaxies {pre-classified} as symmetric $\rm (LG_{A_1}-SG_m)$. The dashed lines represent the median of their respective distribution.
    }
    \label{inc_dist}
\end{figure*}

\subsection{Interpretation of the misclassified cases}

In the previous section we analyzed the results of applying RFs algorithms to the internal parameters of our selected sample of lopsided and symmetric galaxies. In particular, we find that the $\mu_*$ and $T_{\rm P}$ parameters are the primary features used by the classifier to classify galaxies as either lopsided or symmetric, consistent with previous observational studies. However, there are 455 galaxies in the testing set that are misclassified. In this section, we focus on the misclassified cases, $\rm (LG_{A_1}-SG_{m})$ and $\rm (SG_{A_1}-LG_{m})$, to investigate the underlying reasons behind the misclassification.
 
\begin{figure*}[h]
    \centering
    \resizebox{0.75\hsize}{!}
        {\includegraphics{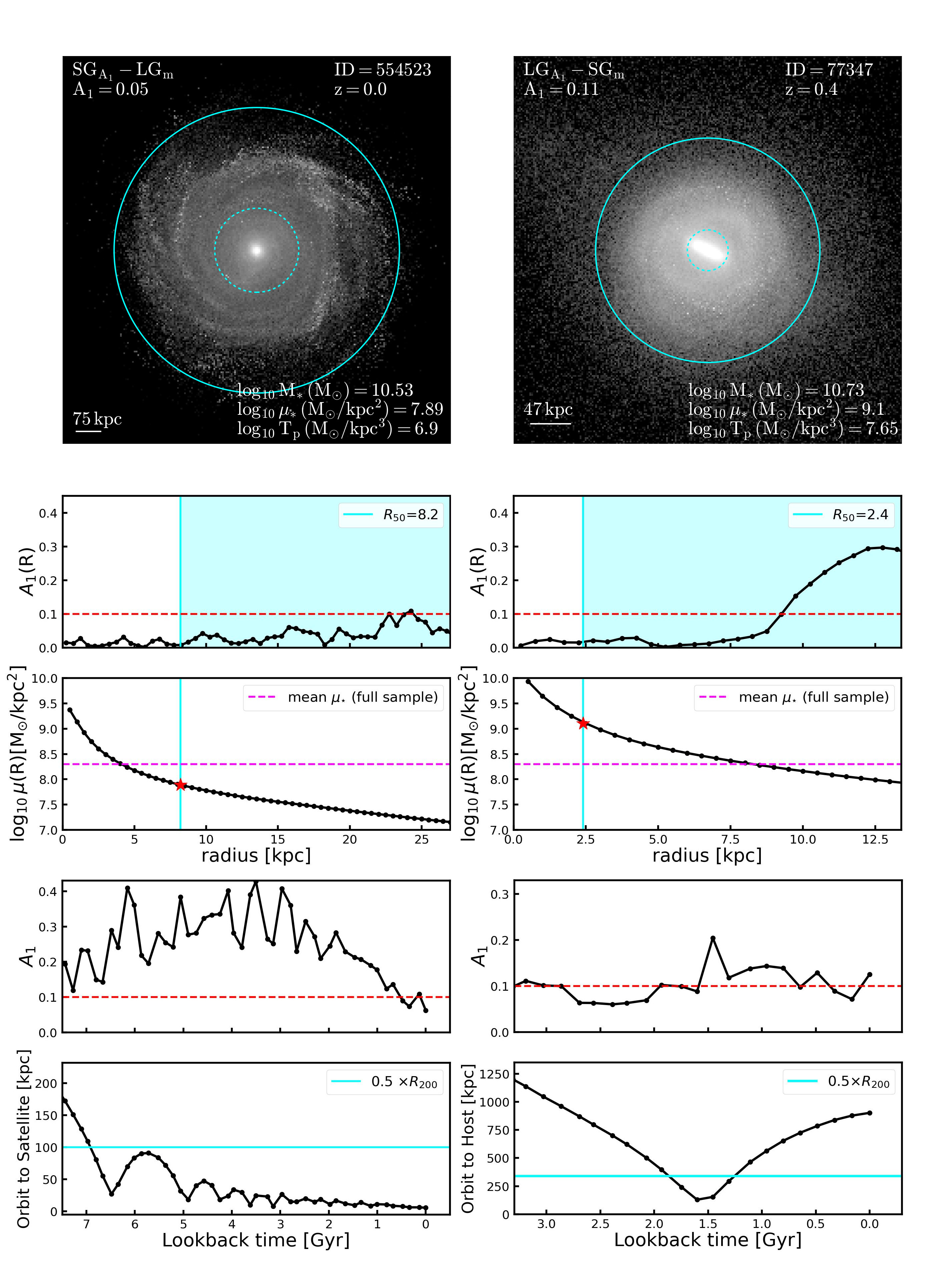}
        }
    \caption{{\textit{(top panels)} V-band face-on projected surface brightness distribution of a \textit{(left)} symmetric galaxy {pre-classified} as lopsided $\rm (SG_{A_1}-LG_m)$ and a \textit{(right)} lopsided galaxy  {pre-classified} as symmetric $\rm (LG_{A_1}-SG_m)$}, considered as examples of the misclassification made by SMOTE+RF. On the upper side, their respective $A_1$ value and classification case are plotted on the left, and their ID and redshift \textit{z} on the right. On the bottom right, the values of the stellar mass ($M_*$), central stellar mass density ($\mu_*$), and tidal parameter ($T_{\rm P}$) are plotted. The dashed cyan lines represent the inner radius $R_{50}$ and the solid cyan lines represent the outer radius 1.4$R_{90}$, which are the limits of the radial interval used in the Fourier decomposition. \textit{(middle panels)} Lopsidedness and stellar density profiles with respect to the radius, up to 1.4$R_{90}$. In both cases, the cyan lines represent the start of the radial interval, $R_{50}$. The  {dashed pink} lines represent the average central stellar mass density ($\mu_*$) of the full sample, with a value of 8.3, while the red stars represent the value of the cental stellar mass density of the galaxy, $\mu_*$, within $R_{50}$. \textit{(Bottom panels)} Lopsidedness and the respective orbit of the most massive satellite with respect to lookback time. The  {dashed red} line represents the $A_1$ threshold to classify lopsided and symmetric galaxies. The horizontal cyan line represents $0.5\times R_{200}$, where $R_{200}$ is defined as the virial radius of the central galaxy.}
    \label{gal-exam2}
\end{figure*}

To further study the incorrectly classified galaxies, in Fig. \ref{a1dist} we highlight their $A_1$ distributions. Again, the  {dashed purple} distribution represents {true lopsided galaxies pre-classified as symmetric galaxies} $\rm (LG_{A_1}-SG_{m})$ with a median of $\sim0.09$. The  {dashed green} distribution represents {true symmetric galaxies pre-classified as lopsided galaxies} $\rm (SG_{A_1}-LG_{m})$ with a median of $\sim0.12$. It is clear that all misclassified galaxies are adjacent to the threshold $A_1=0.1$ and, thus, represent challenging cases for our classification models. In Fig. \ref{inc_dist} we show the distribution of $\mu_{*}$, $T_{\rm P}$, and $R_{\rm ext}$ for all the four different classification cases, following the same color coding as in Fig. \ref{a1vsropt}. Each dashed line represents the median of the corresponding distribution. The top panels show the results obtained from the correctly  {pre-classified} galaxy samples by our model. Note that the distributions differ significantly in all three parameters. As expected, the largest differences are found in $\mu_{*}$. However, even the $R_{\rm ext}$ distributions differ, with median values of 11 $\rm kpc$ and 14.3 $\rm kpc$ for symmetric and lopsided galaxies, respectively. The bottom panels show the distributions obtained from the incorrectly  {pre-classified} samples. Two important things stand out. First, the distributions of the three inspected parameters show more significant overlap with respect to the correctly classified sample. The medians are, in all cases, closer to the median of the overall sample. This is most clear in the $R_{\rm ext}$ distributions, where both symmetric and lopsided nearly perfectly overlap with each other. Second, and most importantly, we find that galaxies classified as lopsided by our global $A_1$ parameter, but {pre-classified} as symmetric by our model $\rm (LG_{A_1}-SG_m)$, have values of $\mu_*$ and $T_{\rm P}$ that are consistent with the distribution of correctly classified symmetric galaxies. In other words, they have relatively large central surface density and $T_{\rm P}$ values. Upon closer inspection of their images, we observe that such galaxies typically display a symmetric overall  {disk}, but a significant asymmetry in their outermost region. An example of  {such a} galaxy is  {shown} in the top right panel of Fig. \ref{gal-exam2}. These localized asymmetries, captured by the global $A_1$ parameter, do not necessarily reflect the overall structure of the  {disk} and can be caused by recent episodes of gas accretion or very recent strong interactions. On the other hand, galaxies classified as symmetric by the global $A_1$ parameter but  {pre-classified as} lopsided by our model $\rm (SG_{A_1}-LG_m)$ show low $\mu_*$ and $T_{\rm P}$ values. Such galaxies display internal properties of typical lopsided galaxies, but simply the morphological perturbation has not yet been triggered. The top left panel of Fig. \ref{gal-exam2} shows an example of  {such a} situation.

To further explore the two examples of misclassified galaxies, in the second and third row of Fig. \ref{gal-exam2} we show their radial $A_1$ and density profiles, respectively. The cyan regions in the second row highlight the radial interval (0.5 - 1.4)$R_{90}$, considered to measure $A_1$. It is worth noting that both galaxies were selected by considering extreme values of $\mu_*$ and $T_{\rm P}$ while having similar stellar mass. For $(\rm SG_{A_1}-LG_m)$, the galaxy shows consistently low $A_1(R)$, even up to the  {disk} outermost regions. Interestingly, its inner stellar density is notably lower than expected for a symmetric galaxy. Even its $\mu_*$, highlighted with a red star, falls below the mean of the overall sample (dashed magenta line). On the other hand, for $\rm (LG_{A_1}-SG_m)$, while $A_{1}(R)$ shows values consistent with 0 within most of the considered radial range, it shows a very strong rise in the  {disk} outskirts. We note that this galaxy has a denser inner stellar region, highlighted by its large $\mu_*$ value which significantly surpass the median of the overall distribution.

To understand these unexpected behavior, we explore on the two lower rows the time evolution of the lopsided parameter and the orbital histories. Interestingly, we find that the $\rm (LG_{A_1}-SG_m)$ galaxy (right panels) became a satellite of a larger host approximately 1.5 Gyr ago. Previous to the pericentric passage, this galaxy showed $A_1$ values below the threshold. After the close interaction, the $A_1$ value rapidly grows as a result of the tidal perturbation of its outer  {disk}. Indeed, we find that this galaxy has internal properties consistent with the symmetric sample, but the strong recent interaction forced an outer tidal disruption, captured by the $A_1$ parameter. In the case of $\rm (SG_{A_1}-LG_m)$ (left panels), the time evolution of $A_1$ shows that, over most of its evolution, this galaxy was indeed strongly lopsided. The initial perturbations was likely induced by significant interaction  with a massive satellite galaxy ($\sim$ 1:10) 6.5 Gyr ago (first pericentric passage). After this point, the galaxy suffered no other interaction with satellite of mass ratios $<$ 1:100. Thus, the lopsided perturbation gradually relaxed, reaching a present-day $A_1$ value below the considered threshold. Even though its internal structure make this galaxy susceptible to lopsided perturbations, the lack of significant external perturbation during its late evolution resulted on a symmetric configuration at the present-day.

\begin{figure*}[h]
 \centering
 \resizebox{0.85\hsize}{!}{
    \includegraphics{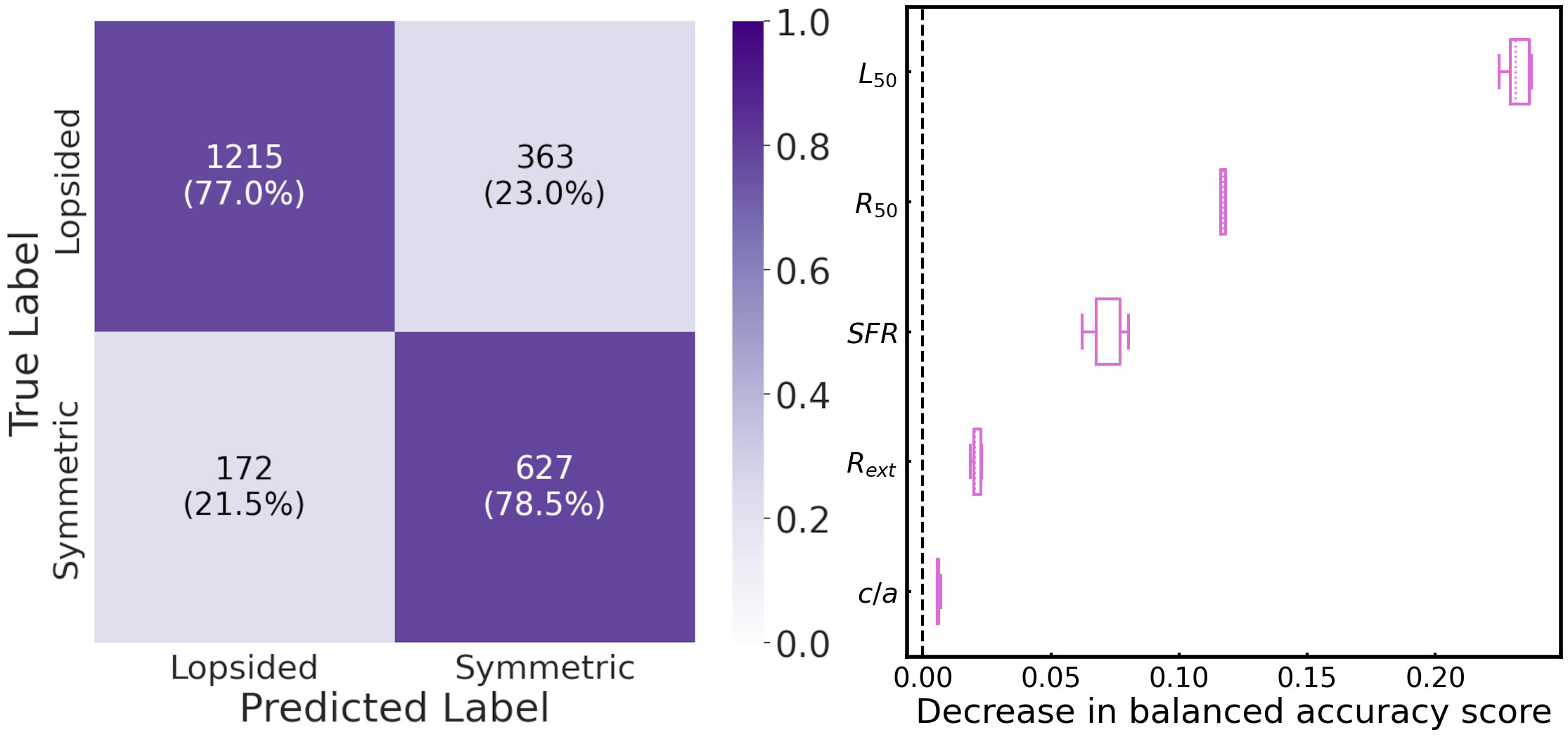}
    }
    \caption{\textit{(left)} Confusion matrix of the testing set using SMOTE+RF with only observational parameters. The  {x axis} is the predicted class or predicted label, and the  {y axis} is the actual class or actual label. The percentage with respect each class is on parenthesis. \textit{(right)} Box plot of each observational feature from the testing set, ranked by their importance as determined by the feature permutation attribute from SMOTE+RF. Each box represents the range of the different scores obtained from a cross-validation with $n_{iter}=5$. The inner dashed line represents the median value of each distribution. The whiskers on each box represent the minimum and maximum value of each distribution.}
    \label{cm_fi}
\end{figure*}

We note that recent interactions cannot explain all the misclassified cases. Indeed, only 76 of the 300 $(\rm LG_{A_1}-LG_m)$ cases are satellite galaxies of a more massive host. Eight (8) additional galaxies have suffered significant interactions ($>$ 1:20) as centrals during the last 3 Gyr. Thus, important interaction can be attributed to this misclassified class in only $28\%$ of the cases. Nonetheless, as previously discussed, other mechanisms such as gas accretion, instability in a counter-rotating  {disk} and torques from an off-centered dark matter halo could be at play in the remaining cases \citep{jog}. Among these mechanisms, asymmetric gas accretion has been proposed as a common driver of lopsidedness. As shown by \cite{bournaud2005}, interactions and mergers can trigger strong lopsidedness in some cases, but they do not account for all the observed statistical properties, such as a correlation between lopsidedness and the Hubble Type, or a correlation between $m=1$ and $m=2$ asymmetries, among others. In a {follow-up study, we shall} focus on the misclassified cases to further the origin of lopsidedness in galaxies with internal properties common to symmetric {disks}.

\subsection{Classification with observable parameters}
\label{subsec_non-corr}
 
Several of the parameters considered in this work require additional modeling to be estimated. Thus, they cannot be directly obtained from observation based on, for example, photometric data. For example, the calculation of $\mu_*$ involves the application of additional stellar population models. Indeed, \cite{reichard2008} calculated the stellar surface mass density following \cite{kauffmann2003} definition, which considers the stellar mass and the Petrosian half-light radius in the z-band. Their stellar massed were estimated using a method that combines spectral diagnostics of star formation histories with photometric data. Additionally, the tidal parameter, $T_{\rm P}$, requires an estimation of the total mass enclosed within $R_{50}$, which involves dynamical modeling of the galaxy.

Despite the importance of these parameters in the classification process, in this section we explore wether it is still possible to obtain a reliable {pre-classification} of lopsided and symmetric galaxies using parameters that are more readily obtainable from photometric data. We follow the same pipeline mentioned earlier, but we train and test the SMOTE+RF classifier with a subset of features that could be estimated from  {multiband} photometric surveys, such as S-Plus \citep{splus} and J-PAS \citep{jpas}. In particular, we replace the parameter $M_{50}$ by the galaxies r-band luminosity  within $R_{50}$, $L_{50}$, thus avoiding the need of stellar population models. In addition to $L_{50}$, we consider as features $R_{50}$, $R_{\rm ext}$, \textit{c/a}, and \textit{SFR}. The later can be obtained from narrow band photometry around the $H_{\alpha}$ line through the Kennicut relation \citep{1998ARA&A..36..189K}. We keep the same hyperparameters listed in Table \ref{hypres}, along with the same training and testing sets. 

The results are presented in Table \ref{results_2}, where it lists the metrics obtained from the testing set. Interestingly, we find very good results, with a performance of the SMOTE+RF algorithm that is only very mildly affected by the limited number of features considered. Indeed, most scores are not significantly affected. Compared to our previous results we find a negligible decrease of $0.4\%$ for balanced accuracy and no change for TNR. Additionally, ROC-AUC has a score of $\sim 80.9\%$, which reflects on how well the model is able to differentiate between both classes. As expected, substituting $M_{50}$ by $L_{50}$ did not introduced a significant drop in the performance. To further characterize our  {pre-classification}, the left panel of Fig. \ref{cm_fi} shows the resulting CM. Note that we obtain a total of 1,927 correctly  {pre-classified} galaxies and only 535 incorrectly  {pre-classified} cases, which represent a $15\%$ increase from the previous  {model}. Compared to our previous results, this model improves in the identification of actual lopsided galaxies, but performs slightly worse in  {pre-}classifying actual symmetric galaxies as symmetric. The feature importance ranking is shown on the right panel of Fig. \ref{cm_fi}, generated with the \textsc{feature\_importance} attribute. We find that the most important parameters are now $L_{50}$, $R_{50}$ and SFR. As before, $c/a$ provides no significant information for the RF classifier.

Our results show that using readily available observational parameters offers a simpler and reliable approach to  {pre-classify} lopsidedness in large observational samples of galaxies, without the need of parameters that required additional modeling to be estimated, such as $\mu_*$ and $T_{\rm P}$. This approach could be particularly valuable in large-scale surveys such as  {the ones that will soon} be provided by LSST \citep{lsst}.

\begin{table}
    \caption{Metric scores of the SMOTE+RF model on the testing set, using only observational parameters.}                
    \centering                          
    \begin{tabular}{l c }      
        \hline              
        \noalign{\smallskip}
        Metric & Score \\
        \noalign{\smallskip} 
        \hline      
        \noalign{\smallskip} 
            Precision& 0.691 $\pm$ 0.012\\
            TPR& 0.770 $\pm$ 0.018\\
            F1-score& 0.728 $\pm$ 0.009\\
            ROC-AUC& 0.799 $\pm$ 0.009\\
            TNR& 0.827 $\pm$ 0.010\\
            G-mean& 0.798 $\pm$ 0.009\\
            Balanced-Accuracy& 0.799 $\pm$ 0.009\\
        \noalign{\smallskip} 
        \hline
    \end{tabular}
    \label{results_2}
\end{table}

\section{Conclusions and discussion}
\label{conc}

In this work we selected a large sample of  {disk}-like galaxies from the IllustrisTNG simulation to develop an algorithm capable of automatically  {pre-classifying} galaxies between lopsided and symmetric. Our main goal was to explore whether this classification can be accurately performed using only internal galactic parameters, thus neglecting information about their present-day environment.

To achieve this we employed the  {random forest} algorithm, a machine learning approach that involves a supervised training process. To label our data as lopsided and symmetric galaxies we employed a Fourier decomposition of the galaxies' stellar density distribution over the radial interval $\rm R_{50}-1.4R_{90}$. We computed a radially average power of the $m=1$ mode, $A_1$, within this range. Galaxies with $A_1 > \rm 0.1$ were classified as lopsided, and the remaining as symmetric. Our sample resulted in a total of 5,273 lopsided and 2,646 symmetric galaxies. To avoid problems in the classification process due to the imbalanced an nature of the dataset, we employed two variations of the RF algorithm: i) we used \textsc{SMOTE} to oversample symmetric galaxies in the training set, thus evening both classes, and ii) we used the BRF algorithm, which balances both classes on each tree by only bootstrapping the minority class while undersampling the majority. Based on the considered metrics, we selected SMOTE+RF as the best model. The classification resulted in a total of 1,922 correctly  {pre-}classified galaxies and 455 incorrectly  {pre-}classified galaxies. This translates in a balanced accuracy of accurate classification rate of $\approx 80\%$ for both classes. To interpret and understand the different decisions leading the RF to the classification, we used a method  {of quantifying ``features importance''}. In particular we utilized an algorithm that randomly permutes the features' values and calculates the decrease in a certain metric; which in our case we choose balanced-accuracy. We found that, to distinguish between both classes, the three most important parameters for the model are $\mu_*$, $T_{\rm P}$, and \textit{SFR}. The excellent results obtained by our classifier, trained with features that do not account for the galaxies environment, strongly supports the hypothesis that lopsidedness is mainly a tracer of galaxies internal structures.

Even though our classifier demonstrated a very good performance, we find that $\approx 20\%$ of the galaxies were misclassified. To study the misclassified cases, we first explore the distribution of the main parameters used by the RF. First, we find that the $A_1$ value of the misclassified cases lies very close to the threshold used to label galaxies as lopsided or symmetric. As a result, these cases are typically associated { with ``borderline classifications''} by $A_1$. Interestingly, we find that the distribution of the most important parameters, such as $\mu_*$ and $T_{\rm p}$ are in good agreement with the class they have been associated  {with} by the RF algorithm. In other words, galaxies classified by $A_1$ as lopsided, but  {pre-classified} as symmetric by the RF, have large $\mu_*$ and $T_{\rm p}$ values. Conversely, galaxies classified by $A_1$ as symmetric but  {pre-classified} as lopsided by the RF, have low $\mu_*$ and $T_{\rm p}$ values. 

To further explore why galaxies with large central surface density and strongly cohesive present perturbed outer  {disk} region, we selected a representative case. We find that the selected galaxy became a satellite of a more massive host $\approx 1.7$ Gyr ago. Previous to the crossing of the host virial radius, the galaxy had a symmetric configuration. However, shortly after its first pericentric passage, its outer regions become perturbed due to the strong tidal interaction.  {Such a} strong and recent interaction induced a temporary lopsided perturbation on this galaxy. We find that $28\%$ of this misclassified class are either satellites of a more massive host, or have had a very recent strong tidal interactions with a massive companion ($>$1:20). For the other misclassified cases, other mechanism, such as asymmetric gas accretion, must be considered to explain the classifications. We  {shall} further explore this in a  {follow-up analysis}. In the case of galaxies with low $\mu_{*}$ and $T_{\rm P}$ that are misclassified as symmetric by the RF algorithm, we find that, typically, they have not experienced recent significant interactions with massive companions. Thus, even though the are susceptible to develop a lopsided perturbation, no external interaction have trigger its onset.

Several parameters considered in this work as features require additional modeling to be estimated. Considering the advent of several surveys such as S-PLUS \citep{splus}, J-PAS \citep{jpas}, and the LSST \citep{lsst}, we explored whether the performance of our classifier significantly drops when considering features that can be readily obtained from multiband photometric surveys. In particular, we replaced stellar mass estimates with their corresponding luminosity in the r-band, and dropped parameters such as $T_{\rm p}$ that involve dynamical modeling to estimate the total galaxy mass within $R_{\rm 50}$. Interestingly, we find the performance of our modeling is very mildly affected, with recovery rates of $\approx 78\%$. These results are very promising, as our algorithm could allow us to rapidly   {preselect} samples of lopsided galaxies from large surveys, allowing us to explore, through other methods such as Fourier decomposition, whether lopsidedness in present-day  {disk} galaxies is connected to their specific evolutionary histories, which shaped their distinct internal properties \citep[][]{arianna}.

\begin{acknowledgements}

 {We thank the anonymous referee again for the valuable comments and suggestions, which have help us improve the final presentation of this manuscript.} V.F. and F.A.G. acknowledge  support from ANID FONDECYT Regular 1211370 and 1251493.  V.F., F.A.G., and A.D. acknowledge  support from the ANID Basal Project FB210003, and the HORIZON-MSCA-2021-SE-01 Research and innovation programme under the Marie Sklodowska-Curie grant agreement number 101086388. M.J.A. acknowledge support from the ANID FONDECYT Iniciaci{\'o}n 11251912.  This article is based on the research for the master’s thesis to obtain the title of MSc. in Astronomy at the University of La Serena. V.F. acknowledges the partial financial support of DIDULS through the project PTE2353855. V.F. also thanks Cristian Vega Martinez for his support on an early version of this project.

\end{acknowledgements}

\bibliographystyle{aa}  
\bibliography{aanda} 

\label{LastPage}
\end{document}